\documentclass[prd,
               twocolumn,
               eqsecnum,
               showpacs,
               letterpaper,
               superscriptaddress,
               altaffilletter,
               nofootinbib, nobibnotes
               ]{revtex4-1}

\usepackage[utf8x]{inputenc}
\usepackage{color}
\usepackage{bm}
\usepackage{graphicx}
\usepackage{amsmath,amssymb}
\usepackage[english]{babel}
\usepackage{amsbsy}
\usepackage{textcomp}
\usepackage{psfrag}
\usepackage{multirow}
 \usepackage{multirow}
 \usepackage{rotating}
\usepackage{sidecap}
\usepackage{verbatim}
\newcommand{\ud}{\mathrm{d}}    
\newcommand{\bwt}{\begin{widetext}}
\newcommand{\ewt}{\end{widetext}}
\newcommand{\beq}{\begin{equation}}
 \newcommand{\eeq}{\end{equation}}
\newcommand{\mycol}{color}
\newcommand{\scale}{0.26}
\begin{document}

\title{Application of asymptotic expansions for maximum likelihood estimators' errors to gravitational waves from inspiraling binary systems: the network case. }
\author{Salvatore Vitale}
\affiliation{Embry-Riddle Aeronautical University, 3700 Willow Creek Road, Prescott, AZ, 86301,USA}
\affiliation{Nikhef - Science Park 105, 1098XG Amsterdam - The Netherlands}
\author{Michele Zanolin}
\affiliation{Embry-Riddle Aeronautical University, 3700 Willow Creek Road, Prescott, AZ, 86301,USA}

\begin{abstract}
This paper describes the most accurate analytical frequentist assessment to date 
of the uncertainties in the estimation of physical parameters from 
gravitational waves generated by non spinning binary systems 
and Earth-based networks of laser interferometers.
The paper quantifies how the accuracy in estimating the intrinsic parameters 
mostly depends on the network signal to noise ratio (SNR), but the resolution in the 
direction of arrival also strongly depends on the network geometry.
We compare results for 6 different existing and possible global networks
and two different choices of the parameter space.
We show how the fraction of the sky where the one sigma angular 
resolution is below 2 square degrees increases about 3 times 
when transitioning from the Hanford (USA), Livingston (USA) and Cascina (Italy)
network to possible 5 sites ones (while keeping the network SNR fixed).
The technique adopted here is an asymptotic expansion of the uncertainties 
in inverse powers of the SNR where the first order 
is the inverse Fisher information matrix.
We show that the commonly employed approach of using a simplified parameter spaces and only the Fisher information matrix can largely underestimate the uncertainties
(the combined effect would lead to a factor ~7 for the one sigma sky uncertainty in square degrees at a network SNR of ~15).
\\
\end{abstract}
\maketitle
\section{Introduction}

With the next generation of advanced ground-based gravitational wave detectors under construction, we analyse the accuracy in the estimation of physical parameters from gravitational waves generated by non spinning binary systems. 
In particular we examine the benefits of proposed extensions of the existing LIGO Virgo detectors network of Hanford (USA), Livingston (USA) (\cite{ALigo}) and Cascina (Italy) (\cite{AVirgo}), to include sites in Australia (\cite{AIGO}), Japan (\cite{LCGT}) and India (\cite{Indigo}).
The paper covers the role of the network in the ability (a) to estimate intrinsic physical parameters and (b) localize sources. This second capability will be critical to aid searches for electromagnetic counterparts to detected gravitational-wave signals.
The analysis is carried out with a novel analytical technique developed in \cite{Zanolin2009} and extends existing studies  (see for example \cite{Zanolin2009}, \cite{Vitale2010}, \cite{Aylott2011} \cite{Vallisneri2008}, \cite{Wen2010}, \cite{Fairhust2010} and the references therein) as discussed below.
The two previous applications of this technique by the authors involved a single optimally oriented interferometer while estimating physical parameters (a) from the inspiral phase of the waveform (\cite{Zanolin2009}) and (b) from the phenomenological inspiral-merger-ringdown waveforms generated by black hole mergers (\cite{Vitale2010}).
In this work we include: (a) the direction of arrival  (b) the orientation of the orbital plane with respect to the observer (c) the use of a network of interferometers and (d) parameter spaces of different sizes. This aspect needs to be addressed carefully since Fisher information matrix can become singular (this is also discussed in see \cite{Vallisneri2008}).\\
\\Recent frequentist analytical results addressing the direction reconstruction, see for example \cite{Wen2010,Fairhust2010} and the references therein, limited themselves to quantify the accuracy in estimating the direction of arrival by using the inverse of the Fisher Information Matrix (IFIM) and a reduced parameter space (the angles were estimated from the times of arrival assuming the intrinsic parameters known). Other articles used the full parameter space, but relied on the  IFIM to calculate the errors \cite{Jaranowski1994,Jaranowski1996}. This can be problematic as we discussed in \cite{Zanolin2009}, \cite{Vitale2010} and here.
In fact, the IFIM delivers accurate error predictions in large signal to noise ratios (SNRs), however for moderate to low SNRs, where the first detections are likely to be recorded, they can grossly underestimate the uncertainties.
This effect is exacerbated when the parameter space is expanded to include intrinsic parameters. The role of the size of the parameter was initially discussed for IFIM in \cite{Arun2005}.
\\In \cite{Zanolin2009},\cite{Vitale2010} we shown how the SNR necessary to trust the inverse of the Fisher information matrix strongly increases when the parameter space is enlarged to include all the unknown parameters in the problem.
We also discussed how at low SNRs the errors (defined as the square root of the sum of the first two order of the expansion of the covariance matrix) decreases like $1/SNR^2$ this is the SNR region where the the inverse of the Fisher information matrix is smaller than higher order contributions we introduced in \cite{Zanolin2009}.
\\In the analysis we observe that increasing the size of the network from 3 to 5 IFOs can increase by a factor 3 the 
fraction of the area that has angular resolution of the order of one or few square degrees.
We also notice that the effects of the size of the parameter space and the second order corrections 
on the direction reconstruction can lead to an increase the uncertainties by about a factor 5 for a network SNR of about 15.
The results of section IV show how the accuracy in the direction reconstruction strongly increases when the network transits from 3 to 4 interferometers but less noticeably with further increases.
On the positive side the details of the analysis show that for regions in the sky where the IFIM predict the smallest uncertainties the impact of the second order and of the size of the parameter space are reduced.

The asymptotic expansions used in this paper can be used to quantify the variance, the bias or the mean square error of the maximum likelihood estimators. When we talk about ”errors” we refer to the square root of the mean square error which is identical to the square root of the variance (or the diagonal elements of the covariance matrix for multivariate estimates) when the bias is zero (\cite{Zanolin2009}). 
Notice that what we call bias is not related to errors in the modelling of the signal or detector, but simply the shift of the mean value of the estimator with respect to the true value of the parameter. The origins of the bias are the non linearity of the estimation process and and sometimes edge effects (if someone estimates a non negative parameter there is usually a positive bias if the true value is close to zero). The results of evaluating the expansions of the bias in the mass bins considered in this paper show that the bias originating from the non linearity of the estimation process is not a concern, leaving only only edge effect biases, when applicable (see \cite{Markowitz2008}).
Finally while the analysis of this paper is performed within a frequentist framework, a parallel effort that did not include India and Japan studied the problem with numerical Bayesian methods (\cite{Aylott2011}). A direct comparison between the two would require to understand the impact of the prior probability distributions and the difference in the questions the two approaches actually try to answer.

In section 2 we review the properties of a received signal and the conventions in defining the direction of arrival and the polarization angle with respect to different reference frames. The expressions are obtained explicitly because none of the many sets of conventions adopted in the literature produced suitable expressions to be inserted in our asymptotic expansions. In the text we provide a more detailed discussion of these issues. 
In section 3 we extend the formalism developed in \cite{Zanolin2009} and \cite{Vitale2010} for a single, optimally oriented interferometer, to a network of interferometers. 
In section 4 we define the waveform and the parameter space. In section 5 we describe the results.
In section 6 we provide some conclusions and recommendations. 
In the appendixes we provide some supporting proofs and mathematical expressions.

\section{The signal on a Earth Based network of detectors}
\subsection{Frames and conventions}

In order to calculate the form of a gravitational wave signal in each detector of a network, it is useful to consider several coordinates frames, as some aspects of the problem may become more clear in a frame than in another. In particular we shall find useful to work in the following frames:

\begin{itemize}
\item {\it Wave Frame (WF).} It has coordinates $(X,Y,Z)$. The GW travels along the positive Z axis; X and Y lie in the directions of the polarization ellipse's axes.
\item {\it Earth Frame (EF).} It has coordinates $(x_E, y_E,z_E)$. The origin is on the center of the Earth, the $x_E$ axis lies in the line joining the center of the Earth with the intersection between the Equator and the Greenwich meridian; $z_E$ passes through the North Pole.
\item {\it I-th-Detector Frame (IDF).} It has coordinates $(x_{I},y_{I},z_{I})$. The axes $x_I$ and $y_I$ lie along the two arms, $\mathbf{n^I_1}$ and $\mathbf{n^I_2}$ respectively. $z_{I}$ points radially out from the Earth surface.
\item {\it Fiducial Frame (FF).} For the moment is a generic frame with respect to whom the positions of the detectors will be specified. Later in the section we identify the fide with the Earth frame. Henceforth we will refer to it as the "fide" frame.
\end{itemize}

We introduce two sets of Euler angles: the first one transforms the fide frame into the wave frame, the second transforms the fide frame into the I-th detector frame:

\begin{eqnarray}
(\phi,\theta,\psi) &\quad, & \quad \mbox{FF} \rightarrow \mbox{WF} \label{EF_to_WF} \\
(\alpha^I, \beta^I, \gamma^I)&\quad, &\quad  \mbox{FF}\rightarrow \mbox{IDF}\label{EF_to_DF}
\end{eqnarray}

Following Goldstein (\cite{Goldstein}), and a big part of the literature, we perform the second rotation around the intermediary x axis (ZXZ convention):

\begin{equation} R(\phi,\theta,\psi) \equiv R_z(\psi)\,R_x(\theta)\, R_z(\phi) \end{equation}

It must be underlined how the Euler angles $\phi$ and $\theta$ are \emph{not} to be identified with the spherical coordinates of the gravitational wave source in the fide frame. 
\\
If one calls $\Phi$ and $\Theta$ the spherical coordinates of the source in the fide frame, $\Xi$ and $\zeta$ the angles that the wave's direction of propagation forms with the fide axes,  the following relations hold:

\begin{equation}\label{Euler_to_Spherical}
\phi= \Phi - \frac{\pi}{2} = \Xi + \frac{\pi}{2} \qquad \theta = \pi -\Theta = \zeta
\end{equation}
these relations, in terms of the latitude lat. and longitude long. of the source in the fide frame, become:

\begin{equation}\label{Euler_to_Geo}
\phi= long - \frac{\pi}{2} = \Xi + \frac{\pi}{2} \qquad \theta = \frac{\pi}{2} +lat = \zeta.
\end{equation}

One can find similar relations between longitude $\Omega^I$ and latitude $\Upsilon^I$ of the I-th detector beam splitter and the Euler angles $\alpha^I,\beta^I$. The third Euler angle, $\gamma^I$ can be fixed noticing that after the second rotation the intermediate frame will have its x axis along the local parallel, pointing East, and the y axis along the local meridian, pointing North. The third Euler's rotation must align the axes with the arms. We label the arms in such a way that the first arm has only the x component, while the second has only the y component. If we call $\Delta^I$ the angle between the first arm and the local North direction, then:

\begin{equation}\label{Euler_to_Latitude}
\alpha^I = \Omega^I+\frac{\pi}{2}\;,\; \beta^I= \frac{\pi}{2}-\Upsilon^I\;,\; \gamma^I= \Delta^I+\frac{\pi}{2} 
\end{equation}

The local Earth radius at the position of the I-th detector's beam splitter is indicated with $r_I$.

In the following sections, necessary different usages of indexes are introduced as below:

\begin{enumerate}
\item Tensor indexes.  Lower case Latin letter, usually $(i,j,k,..)$, will indicate the components of tensors when the range is $1..3$ (Greek letter if the range is $0..3$). The Einstein convention for these indexes is always used, unless the contrary is explicitly stated.
\item When a quantity refers to the I-th interferometer, we add $I$ on the upper right side of the symbol. E.g. $d^{I}$
\item When we calculate explicitly the components of a vector or tensor in a particular frame, the frame where the expression holds is indicated with an index in the upper left of the symbol. E.g. ${\,}^{\mbox{IDF}}\!d^{I}$ will be used to give the components of the I-th detector tensor in the I-th detector frame.
\end{enumerate}

\subsection{Interferometer output}\label{SSection_IFOOutput}

Gravitational waves are \emph{ripples} in the space-time, that can be represented as fluctuations $h_{\mu\nu}$ around the background flat metric. In the transverse-traceless (TT) frame (\cite{MTW}, \cite{Thorne300}) the only non vanishing components of $h_{\mu\nu}$, if the wave is traveling in the positive $Z$ direction, are $h_{XX}=-h_{YY}\equiv h_+$ and $h_{XY}=h_{YX}\equiv h_\times $ \footnote{Some authors, notably \cite{Schutz1987} and \cite{Dhurandhar1988}, put an explicit imaginary unit in the cross polarization, then writing $h_{XY}=h_{YX}\equiv i h_\times $\label{foot1}}, where $h_+$ and $h_\times$ are the two independent polarizations. 

While dealing with gravitational waves, it is useful to introduce a $3\times3$ \emph{wave tensor}, defined as one half of the spatial part of the metric perturbation: $$ w_{i\,j}\equiv \frac{1}{2}\, h_{i\,j}\qquad i,j=1..3 .$$

In an arbitrary frame, the wave tensor can be expressed in terms of its circular polarization components (\cite{Pai2000}):

\begin{equation}\label{wave}
 w^{ij}(t) = \frac{1}{2} \left[ (h_+(t)+i h_\times(t) ) e^{ij}_{R} + (h_+(t)-i h_\times(t) ) e^{ij}_{L}\right]
\end{equation}

where $e^{ij}_{R,L}$ are the right and left circular polarization tensors unity. 
They are orthonormal one with respect to the other:

\begin{equation}
 e^{ij}_R e_{R\;ij}^{*} = 1 ,\qquad e^{ij}_R e_{L\;ij}^{*} =0 ,\qquad e^{ij}_L e_{L\;ij}^{*} = 1 
\end{equation}

Since the components of the wave tensor are real:

\begin{equation}\label{base_reality}
 e^{ij}_R=e^{ij\;*}_L
\end{equation}

In any reference frame, the polarization tensors can be expressed as:

\begin{equation}\label{base_as_m}
 e^{ij}_L= m^i m^j
\end{equation}

where $m^i$ are the components of the null vector:
\begin{equation}
m^i= \frac{1}{\sqrt{2}} \left(e^i_X + i e^i_Y\right)
\end{equation}

with $\mathbf{e}_X$ and $\mathbf{e}_Y$ unit vectors along the WF axes.
Using (\ref{base_reality}) and (\ref{base_as_m}), the wave tensor (\ref{wave}) can be written in an alternative way, that will be useful later:

\begin{equation}\label{wave_generic}
 w^{ij}(t)= h_{+}(t)\, \Re[ m_i\,m_j] + h_{\times}(t) \, \Im[m_i\,m_j]
\end{equation}

In the wave the null vector $\mathbf{m}$ has the simple form:

\begin{equation}\label{null_vector_standard}
{\,}^{\mbox{WF}}\!m^i= \frac{1}{\sqrt{2}} (1,i,0)
\end{equation}

and the wave tensor has the explicit expression:
  
\begin{equation}\label{wave_in_WF}
 {\,}^{\mbox{WF}}\! w= \frac{1}{2} \left( \begin{array}{ccc}
 h_{+}&h_\times&0\\
 h_\times& - h_{+} &0\\
0&0&0
  \end{array}\right)
\end{equation}

The form of the wave tensor in another frame is easily calculated expressing the components of the null vector $\mathbf{m}$ in that frame and using eq. (\ref{wave_generic}).
The GW signal at the I-th detector can be written as:

\begin{equation}\label{Signal_I}
s^I(t)= w^{ij}\left[t - \tau_I(\Xi^I,\zeta^I)\right] d^I_{ij}
\end{equation}

where 
 $d^I_{ij}$ is the I-th \emph{detector tensor}  (\cite{Schutz2009}). For detectors with orthogonal arms, like those we consider in this work, $\mathbf{d}$ has the form:

\begin{equation}\label{Detector_Tensor}
d^I_{ij}= n^I_{1\,i}\,n^I_{1\,j} - n^I_{2\,i}\,n^I_{2\,j}.
\end{equation}

$\mathbf{n^I}_{1}$ and $\mathbf{n^I}_{2}$ are the unit vector along the I-th interferometer arms. 

In eq. (\ref{Signal_I}), the wave tensor is evaluated at the retarded time $t - \tau_{I}$ that takes into account the time the GW needs to travel from the I-th detector site to the fide frame:

\begin{equation}\label{time_shift}
\tau_{I} \equiv t_{IDF} - t_{FF} = \frac{\hat{\mathbf{n}} \cdotp (\mathbf{r_I}-\mathbf{r_F})}{c}
\end{equation}

In the previous equation, $\mathbf{r_I}$ and $\mathbf{r_F}$ are the positions of the I-th detector's beam splitter and the fide frame origin in an arbitrary frame, while $\hat {\mathbf{n}}$ is a unit vector in the direction of the wave propagation, i.e. on the $Z$ axis.
The time shift is a scalar, and can be calculated in any frame. However, the calculations are simpler in the fide frame, where $\mathbf{r_F}=0$ and 

\begin{eqnarray}
{\,}^{\mbox{FF}}\!\mathbf{n}(\Xi,\zeta)&=& (\sin\zeta\cos\Xi, \sin\zeta\sin\Xi,\cos\zeta)\\
{\,}^{\mbox{FF}}\!\mathbf{r_I}(r_I,\Omega^I,\Upsilon^I)&=& r_I \,(\cos\Upsilon^I\cos\Omega^I, \sin\Upsilon^I\cos\Omega^I,\sin\Upsilon^I)\nonumber
\end{eqnarray}

(the relations between these angles and Euler's were given before, eqs. (\ref{Euler_to_Spherical}) and (\ref{Euler_to_Latitude})).

From eq. (\ref{Detector_Tensor}) it follows that each detector tensor, $\mathbf{d}^I$ has a very simple expression in its own frame:

\begin{equation}\label{Detector_tensor2}
{\,}^{\mbox{IDF}}\!d^{I}= \left( \begin{array}{ccc}
 1&0&0\\
0& -1	 &0\\
0&0&0
  \end{array}\right)
\end{equation}

If we introduce the null vector:

\begin{equation}\label{null_Ith}
{\boldsymbol \rho}^I\equiv \sqrt{\frac{1}{2}} \left( \mathbf{e_{x{_I}}} + i \;\mathbf{e_{y_I}}\right) 
\end{equation}

where $\mathbf{e_{x{_I}}}$ and $\mathbf{e_{y{_I}}}$ are the unit vectors along the IDF axes, then the detector tensor can be written in any frame as:

\begin{equation}\label{Detector_tensor_generic}
 d^I_{ij}= 2 \Re [\rho^I_i\,\rho^I_j]
\end{equation}

just expressing the components of ${\boldsymbol \rho}^I$ in the frame of interest.
\\
The signal (\ref{Signal_I}) can be calculated explicitly in any frame. Being an invariant, the final result will be the same, but the calculations can be somehow simpler in a particular frame. 

Let us consider the wave frame, for example, in which the wave tensor has the simple form (\ref{wave_in_WF}). We need to write the signal at the I-th detector in the WF, which can be done calculating the components of the null vector $\boldsymbol{\rho}^I$ in the WF:

$$ {\,}^{\mbox{WF}}\!\rho^I_{i\,j} = R(\phi,\theta,\psi)_{i\,p}\;R^{-1}(\alpha^I,\beta^I,\gamma^I)_{p\,q} {\,}^{\mbox{IDF}}\!\rho^I_{q\,j}$$

and using eq. (\ref{Detector_tensor_generic}).
The signal acquires the form:
\begin{widetext}
\begin{equation}\label{signal_I_multi}
s^I(t) = h_{+}(t-\tau_{I})\Re \left[ {\,}^{\mbox{WF}}\rho_{11}{\,}^{\mbox{WF}}\rho_{11}- {\,}^{\mbox{WF}}\rho_{22}{\,}^{\mbox{WF}}\rho_{22} \right]+ 2\,\,h_{\times}(t-\tau_{I}) \Re[ {\,}^{\mbox{WF}}\rho_{12}{\,}^{\mbox{WF}}\rho_{12}]
\end{equation}
\end{widetext}

We do not derive here the explicit value, as it is rather cumbersome. 

Given the general formula, (\ref{signal_I_multi}) we can recover the single detector result in the following way: suppose the I-th detector is the only one present, so that we can identify its frame with the fide frame. This implies that the angles $(\alpha^I,\beta^I,\gamma^I)$ and the time delay $\tau_I$ are zero, and that the matrix $R^{-1}(\alpha^I,\beta^I,\gamma^I)$ is the identity matrix.
With some simple algebra is possible to show that in this case the signal (\ref{signal_I_multi}) can be written as follows:

\begin{equation}\label{signal_I_single}
s(t) = h_{+}(t) F_+ + \,\,h_{\times}(t) F_{\times}
\end{equation}

where the \emph{antenna pattern} have the following explicit expressions in terms of the Euler angles:
\begin{widetext}
\begin{eqnarray}
 F_{+}&=& \frac{1}{2} (\cos^2\theta +1) \cos2\phi\cos 2\psi - \cos\theta \sin2\phi\sin2\psi \label{antenna_pattern_p}\\
F_\times &=& -\frac{1}{2}(\cos^2\theta +1) \cos2\phi\sin 2\psi - \cos\theta \sin2\phi\cos2\psi \label{antenna_pattern_c}
\end{eqnarray}
\end{widetext}

The expressions obtained are not immediately comparable with other present in the literature, due to the many different conventions used in the years. Let us just compare our result with some notable articles.
Our antenna patterns become the same of \cite{Schutz2009} if we express them in terms of $\Xi$ and $\zeta$ (eq. \ref{Euler_to_Spherical}), because \cite{Schutz2009} writes them as functions of the angles between the propagation vector and the fide axes.
More attention is required in order to compare our result with those in \cite{Schutz1987} and \cite{Dhurandhar1988}: in these articles the x axis is not along one of the detector's arms, but instead it bisects the arms. This introduce a $\pi/4$ shift in the definition of the first Euler's angle. But there is more: in the cited paper the first arm is in the positive xy quadrant, while the second arm has a negative y component. 
This means that a relabel $n_1 \leftrightarrow n_2$ of the arms is also required. This operation changes the sign of the detector tensor, and then of the antenna patterns. Multiplying eqs. \ref{antenna_pattern_p} and \ref{antenna_pattern_c} by $-1$, and adding a $\pi/4$ to $\phi$ one proofs that our antenna patterns are equal to those in the references (an extra imaginary unit must be added in the cross polarization, see footnote \ref{foot1} at the beginning of this paragraph)
  
\subsection{Expansion on a Symmetric-Trace-Free base}

In \cite{Schutz1987}, \cite{Dhurandhar1988} and \cite{Pai2000}, the problem of the signal form at the detector is treated in an elegant way, recognizing that both the detector and the wave tensors are symmetric and trace-free (STF) tensors, and can be developed in a base of STF tensors.

We have seen that the wave tensor can be expressed in any frame in terms of the real and imaginary part of the matrix $\mathbf{m}\otimes \mathbf{m}$, (the matrix whose $(i,j)$ element is $m_i m_j$).
In the wave frame $\mathbf{m}$ has the components given in eq. (\ref{null_vector_standard}), while in a different frame, say the FF, which can be obtained from the WF with rigid rotation, its components will depend on the Euler angles that describe the rotation. More precisely, it can be shown that the real and the imaginary parts of the matrix $\mathbf{m}\otimes \mathbf{m}$ in the FF are:

\begin{equation}\label{Re_generic}
{\,}^{\mbox{FF}}\! \Re [ m_i  m_j]= \sqrt{\frac{2\pi}{15}} \left[T_{2n}(\phi,\theta,\psi) + T_{-2\,n}(\phi,\theta,\psi)\right] \mathcal{Y}^n_{i\,j} \end{equation}
\begin{equation}
{\,}^{\mbox{FF}}\! \Im [ m_i  m_j]= -i \sqrt{\frac{2\pi}{15}} \left[T_{2n}(\phi,\theta,\psi) - T_{-2\,n}(\phi,\theta,\psi)\right] \mathcal{Y}^n_{i\,j} \label{Im_generic}
\end{equation}
 
Where the functions $T_{mn}$ are the second-order \emph{Gel'fand functions}, and $\mathcal{Y}^n_{i\,j}$ are a base for rank 2 STF tensors.
Both these quantities, together with some useful mathematical background, will be discussed in appendix \ref{Appendix_Gelfand}.

As told before, both the wave tensor and the I-th detector tensor are STF, and can thus be written in terms of the Gel'fand functions, using (\ref{wave_generic}), (\ref{Detector_tensor_generic}), (\ref{Re_generic}) and (\ref{Im_generic}) with the appropriate rotation angles.

Let us calculate, for example, the signal in the wave frame, were the wave tensor is (\ref{wave_in_WF}) and the I-th detector tensor has the form:
\begin{widetext}
\begin{equation}\label{detector_in_WF_STF}
{\,}^{\mbox{WF}}\!d^I_{ij}= 2 \Re[\rho^I_I \rho^I_j]= 2 \sqrt{\frac{2\pi}{15}} \big[{T_{2 \,n}}(\mbox{WF} \rightarrow \mbox{IDF}) + {T_{-2\,n}}(\mbox{WF} \rightarrow \mbox{IDF})\big] \mathcal{Y}^n_{ij}\equiv  2 \sqrt{\frac{2\pi}{15}} \chi_n\, \mathcal{Y}^n_{ij}
\end{equation}
 \end{widetext}

where we have indicated with the syntax $F(\mbox{WF} \rightarrow \mbox{IDF}$ the fact that the function $F$ depends on the Euler angles that rotate WF into IDF, and where we have introduced the combination:

\begin{equation}\label{chi}
\chi_n\equiv {T_{2 \,n}}(\mbox{WF} \rightarrow \mbox{IDF}) + {T_{-2\,n}}(\mbox{WF} \rightarrow \mbox{IDF}).
\end{equation}

Using the symmetry (\ref{simmetry_T}) of the Gel'fand functions, it is simple to verify that
 \begin{equation}
\chi_{-n}= \chi_n^* .
\end{equation}

This allows us to write the detector tensor in the WF in the simpler form:

\begin{equation}\label{detector_in_WF_chi}
{\,}^{\mbox{WF}}\!d_{ij}= \left(\begin{array}{ccc}
    \Re [\chi_2]& -\Im [\chi_2]&0\\ - \Im[\chi_2]& -\Re[\chi_2]&0\\0&0&0
            \end{array}\right)
\end{equation}

Multiplying (\ref{wave_in_WF}) by (\ref{detector_in_WF_chi}) the signal at the I-th detector takes the form:

\begin{equation}
 s^I(t)= h_+ \Re[\chi_2] - h_\times \Im[\chi_2] = \Re[h\, \chi_2]
\end{equation}

with $h\equiv h_+ +i h_\times$.

This expression looks beautifully compact, but it is not very useful for real calculations. Let us write down the explicit value of $\chi_2$ in terms of Gel'fand functions:

\begin{eqnarray}
 \chi_2&=& T_{22}(\mbox{WF}\rightarrow \mbox{IDF}) + T_{-22}(\mbox{WF}\rightarrow \mbox{IDF})=\nonumber\\
&=& T_{2s}(\mbox{FF}\rightarrow \mbox{IDF})\,T_{s2}(\mbox{WF} \rightarrow \mbox{FF}) +\nonumber \\ 
&+& T_{-2s}(\mbox{FF}\rightarrow \mbox{IDF})\,T_{s2}(\mbox{WF} \rightarrow \mbox{FF}) \nonumber\\
&=& T_{s2}(\mbox{WF}\rightarrow \mbox{FF})\left[T_{2s}(\alpha^I,\beta^I,\gamma^I) + T_{-2s}(\alpha^I,\beta^I,\gamma^I)\right]\nonumber\\
&=& T^*_{2s}(\phi,\theta,\psi)\left[T_{2s}(\alpha^I,\beta^I,\gamma^I) + T_{-2s}(\alpha^I,\beta^I,\gamma^I)\right]\label{chi_explicit}
\end{eqnarray}

where we have used the addition formula (\ref{Gelfand_sum}) for Gel'fand functions, while passing from first to second line, and the relation (\ref{Gelfand_inverse}) in the last line, to write the Gel'fand functions associated with the inverse rotation $\mbox{FF}\rightarrow \mbox{WF}$.

The real an imaginary part of $\chi_2$ can be then written:
\begin{widetext}
\begin{eqnarray}
 \mathcal{F}_+^I\equiv\Re[\chi_2]&=& \frac{1}{2} \left(T_{2s}(\alpha^I,\beta^I,\gamma^I)+ T_{-2s}(\alpha^I,\beta^I,\gamma^I)\right)\left(T_{2s}^*(\phi,\theta,\psi)+T_{-2s}^*(\phi,\theta,\psi)\right) \label{Generalized_antenna_p}\\
 \mathcal{F}_\times^I\equiv - \Im[\chi_2]&=& \frac{i}{2} \left(T_{2s}(\alpha^I,\beta^I,\gamma^I)+ T_{-2s}(\alpha^I,\beta^I,\gamma^I)\right)\left(T_{2s}^*(\phi,\theta,\psi)-T_{-2s}^*(\phi,\theta,\psi)\right)\label{Generalized_antenna_c}
\end{eqnarray}
\end{widetext}

where we have introduced the generalized antenna pattern $\mathcal{F}_+$ and $\mathcal{F}_\times$, using which the signal at the I-th detector looks formally as in the single detector case:

\begin{equation}\label{signal_final}
s^I(t)= h_+(t-\tau_I)\, \mathcal{F}_+ +  h_\times(t-\tau_I) \,\mathcal{F}_\times
\end{equation}

We won't develop these expressions, because of their size.
Instead we can check that the generalized antenna patterns have the single IFO limit values (\ref{antenna_pattern_p} and \ref{antenna_pattern_c}) when a single detector is present in the Network, this is done in appendix \ref{Appendix_Generalized_to_Single}.
The expression we have obtained here is compatible with \cite{Pai2000} once the differences in the definitions of the detector frames are taken into account, which results in a $\pi/4$ shift in the definition of $\gamma^I$.

\section{Fisher Matrix, CRLB and higher orders}

We write the output of the I-th detector as the sum of the signal $s^I(t)$ and the noise $n^I(t)$:

\begin{equation}
 x^I(t) = s^I(t) + n^I(t)
\end{equation}

The signal $s^I$ will generally depend on a vector of unknown parameters $\bm{\vartheta}$ which we want to estimate, that can be the physical parameters of the source (e.g. total mass) as well as extrinsic parameters as the position on the sky (see section \ref{Section_WF}). 
However, here and in what follows, to unburden the expressions, we do not write explicitly this dependence, as well as the dependence on the time delay $\tau_I$, by using the notation:

\beq
s^I(t)\equiv s^I(t-\tau_I; \bm{\vartheta}).
\eeq

If we assume that the noise in each detector is stationary and Gaussian with zero mean, the probability distribution for the data realization in the I-th interferometer is given by:
\bwt
\begin{equation}\label{Probability_I}
 p(x^I) \propto \exp{\left\{-\int dt dt' (x^I(t)-s^I(t))\Omega^I(t-t')(x^I(t')-s^I(t')) \right\}}
\end{equation}
\ewt
where $\Omega^I(t-t_1)$ is the inverse of the noise correlation matrix for the I-th detector. 
The noise from laser interferometers is usually a combination of smaller Gaussian fluctuations and larger, rarer non Gaussian outliers (``glitches'' in the data).
The use of coincident requirements between different sites and a whole set of data quality and vetoes procedures make it reasonable the assumption that glitches will be recognized an removed or vetoed from the data, leaving only 
Gaussian distortions to the GW signal. We also assume that the noise in different IFOs is statistically independent.
In this case the joint probability distribution is simply the product of (\ref{Probability_I}) for $I=1..N$:

\begin{equation}\label{Probability_Tot}
 P(\mathbf{x})= \prod_{I=1}^N p(x^I)
\end{equation}

and the log-likelihood turns out to be the sum of the log-likelihood of the detectors:

\begin{equation}\label{Loglikelihood}
 \ln[P(\mathbf{x})]\equiv \ell= \sum_{I=1}^N \ln[p(x^I)] =\sum_{I=1}^N \ell^I
\end{equation}

The additivity will hold for functions that are built from (\ref{Loglikelihood}) through linear operations.
This is true in particular for the Fisher information matrix defined as:

\begin{equation}\label{FisherNetw}
 \Gamma_{i\,j}= -E\left[\ell_{ij}\right] = -\sum_{I=1}^N \mathcal{E}^I\left[\ell^I_{ij}\right]= \sum_{I=1}^N \Gamma^I_{i\,j}
\end{equation}

where $E[\cdotp]$ is the expectation with the joint pdf (\ref{Probability_Tot}), $\mathcal{E}^I[\cdot]$ is the expectation with the I-th marginal probability and where we have defined:

\begin{equation}
 \ell_{i \cdots j} \equiv \frac{\partial^{(n)} \ell}{\partial \vartheta^i \cdots \partial \vartheta^j}
\end{equation}

The same notation will be used for the signal derivatives with respect to the components of the vector of unknown parameters $\bm{\vartheta}$, e.g. $s_i \equiv \frac{\partial s}{\partial \vartheta^i}$

The single detector Fisher matrix $\Gamma^I_{ij}$ can be expressed in the Fourier domain and it is possible to prove (see for example \cite{Zanolin2009}) that the final expression is:

\begin{eqnarray}\label{Fourier_Mean}
 \Gamma^I_{ij} &=& 4\,\Re \int_{f_{low}}^{f_{up}}{\ud f \frac{s^I_i(f) s^{I*}_j(f)}{S^I(f)}}
\end{eqnarray}

where we have introduced the I-th detector {\em one-sided noise spectral density} $S^I(f)$ (\cite{Maggiore}, \cite{Schutz2009}). 
The lower cut-off frequency, $f_{low}$ is a detector dependent quantity that we fix when the noise models for the detectors are introduced. The upper limit of integration $f_{up}$, which is waveform dependent, indicates up to which frequency we are confident that the waveform we use is correct. These limits will be set later in the section.

Noise-weighted integral like this one, between derivatives of the signal will, are denote with wedge brackets:

\begin{eqnarray}
 \langle s_{a \cdots i} | s_{j \cdots  p} \rangle &\equiv& \langle a \cdots i | j \cdots p \rangle= \langle \frac{\partial s^{(n)}}{\partial \theta^a \cdots \partial \theta^i} | \frac{\partial s^{(m)}}{\partial \theta^j \cdots \partial \theta^p}  \rangle \nonumber \\
&=& 4 \Re \int_{f_{low}}^{f_{cut}}{ \ud f\, \frac{s_{a \cdots i}s_{j \cdots  p}^* }{S(f)}}\end{eqnarray}

The optimal SNR, $\rho^I$, of the signal at the I-th detector can also be written in a similar way  (\cite{Schutz2009}) :

\beq\label{snr_Ith}
(\rho^I)^{\,2} \equiv 4\,\Re \int_{f_{low}}^{f_{up}}{\ud f \frac{s^I(f) s^{I*}(f)}{S^I(f)}} = \langle s(f)|s(f)\rangle^I
\eeq

It is intuitive that adding interferometers to our network, while keeping everything else fixed, will increase the amount of information we have on a signal that is present. We can take this into account introducing the \emph{network SNR}, $\rho$ defined as:
\beq\label{snr_net}
\rho^2 \equiv \sum_{I=1}^{N}(\rho^{I})^2
\eeq

which is the SNR we quote in the results section.
Once the Fisher matrix is calculated, its inverse can be used to estimate errors and covariances.

The Cramer Rao lower bound (CRLB) gives a bound for the variance of the i-th parameter (\cite{Rao}):

\begin{equation}\label{CRLB}
(\Delta\vartheta^i)^2 \geq \left[\Gamma^{-1}\right]_{ii} \equiv \Gamma^{ii}
\end{equation}

while the covariance between the i-th and the j-th parameter is:

\begin{equation}\label{Covariance}
Cov(\vartheta^i,\vartheta^j) \geq \left[\Gamma^{-1}\right]_{ij} \equiv \Gamma^{ij}.
\end{equation}

In a high SNR regime, the errors and covariances are close to these bounds, and one writes the probability distribution for the errors as a multivariate Gaussian:

\begin{equation}
p(\Delta \bm{\vartheta}) \propto e^{-\frac{1}{2} \Delta\vartheta_i \Delta\vartheta_j \Gamma_{i\,j}}
\end{equation}

with $i,j = 1.. M$, $M$ being the dimension of $\bm{\vartheta}$.

On the other hand, the CRLB is known to fail for small SNRs, that is, for weak signals. In those situations numerical simulations, like MonteCarlo simulations (MC) are more faithful, but at the price of an high computational cost.
In \cite{Zanolin2009} we proposed an analytical method to improve the errors estimation for small SNRs, and we used it for a 3.5PN signal. In \cite{Vitale2010} we applied the same method to an Inspiral-Merger-Ringdown (IMR) signal (\cite{Ajith2007,Ajith2007b,Ajith2009}), also showing that the bias arising from the nonlinearity of the estimator plays a fundamental role for high mass systems.

In \cite{Zanolin2009} and \cite{Vitale2010} we have shown how the bias and (co)variance of the estimators can be written as power series in one over the SNR $\rho$:

\begin{eqnarray}
\sigma_{\vartheta^i \vartheta^j}^2 &=& \frac{S_{ij}^2[1]}{\rho^2} +\frac{S_{ij}^2[2]}{\rho^4}+\cdots= \sigma^2_{\vartheta^{i}\vartheta^{j}}[1]+\sigma^2_{\vartheta^{i}\vartheta^{j}}[2] +\cdots \label{Sigma_series}\\ \nonumber \\
b_{\vartheta^i} &=& \frac{B_i[1]}{\rho} + \frac{B_i[2]}{\rho^2} + \cdots=b_{\vartheta^i}[1] +b_{\vartheta^i}[2] + \cdots \label{bias_series}
\end{eqnarray}

where after the first equal sign we have shown explicitly how the different terms depend on the SNR. 
We have also shown that the first order in the variance series is the usual CRLB, while the second orders contain higher derivatives of the signal. 
Formally, we can still use the expressions we gave there, for example for the second order diagonal elements of the covariance matrix:

\begin{widetext}
\begin{eqnarray}\label{VarMatrix}
\sigma^2_{\vartheta^i\vartheta^i}[2]\equiv \sigma^2_{\vartheta^i}[2]\!&=&\!-\Gamma^{jj}
\!+\!\Gamma^{jm} \Gamma^{jn} \Gamma^{pq}(2\upsilon_{nq,m,p}\!+\!\upsilon_{nmpq}\!+\!3\upsilon_{nq,pm}\!+\!2\upsilon_{nmp,q}\!+\!
\upsilon_{mpq,n})\!+\nonumber\\
&+&\Gamma^{jm}\Gamma^{jn}\Gamma^{pz}\Gamma^{qt}\bigg[(\upsilon_{npm}\!+\upsilon_{n,mp})(\upsilon_{qzt}+2\upsilon_{t,zq})+\upsilon_{npq}\left(\frac{5}{2}\upsilon_{mzt}+2\upsilon_{m,tz}+\upsilon_{m,t,z}\right)\nonumber\\
&+&\upsilon_{nq,z}(6\upsilon_{mpt}+2\upsilon_{pt,m}+\upsilon_{mp,t})\bigg]
\end{eqnarray}
\end{widetext}

but now both the network Fisher matrix inverse $\Gamma^{ i j}$ and the $\upsilon$ must be calculate using the joint network likelihood.

The second corrections contain the factor:

\begin{equation*}
{\upsilon_{a_1 a_2..a_s ,\,\,..\,\,, b_1 b_2..b_s}=
E\left[ \ell_{a_1 a_2.. a_s}\,\,..\,\,\ell_{b_1 b_2..b_s}\right] }. 
\end{equation*}

where now $\ell$ is the joint log likelihood, eq. (\ref{Loglikelihood}). It is easy to check that it is not possible to just add up the single-IFO contributions, as ``cross terms'' between different interferometers might appear.

Let's consider for example:

\begin{eqnarray}\label{UpsilonABCD}
\upsilon_{ab,cd} &=& E\left[\ell_{ab} \ell_{cd}\right]= \sum_{I}^N\sum_{J}^N E\left[ \ell^I_{ab}\ell^J_{cd}\right].
\end{eqnarray}

The terms with $I=J$ will give back the sum of single-IF0 $\upsilon_{ab,cd}^I$, but the terms with $I\neq J$ also give a contribution, for example for the $(1-2)$ term (note the apex at the end of the square brackets to distinguish the contributions of IFO 1 and IFO 2):

\begin{widetext}
\begin{eqnarray}
 E\left[ \ell^1_{ab}\ell^2_{cd}\right]&=& \int{\ud t \ud t' \ud k \ud k' \ud \mathbf{x} P(\mathbf{x})} \left[s_{ab}(t,\vartheta) \Omega(t-t') \bigg(x(t')-s(t',\vartheta)\bigg) - s_{a}(t,\vartheta) \Omega(t-t') s_b(t',\vartheta)  \right]^{(1)}\times\nonumber\\
&\times& \left[s_{cd}(s,\vartheta) \Omega(k-k') \bigg(x(k')-s(k',\vartheta)\bigg) - s_{c}(k,\vartheta) \Omega(k-k') s_d(k',\vartheta)  \right]^{(2)}  =\nonumber\\
&=& \int{\ud t \ud t'  \ud x^1  p(x^1) s^{(1)}_{a}(t,\vartheta) \Omega^{(1)}(t-t') s^{(1)}_b(t',\vartheta)}\int{\ud k \ud k'\ud x^2 p(x^2)  s^{(2)}_{c}(k,\vartheta) \Omega^{(2)}(k-k') s^{(2)}_d(k',\vartheta)}\nonumber\\
&=& \Gamma^1_{ab} \Gamma^2_{cd}
\end{eqnarray}
\end{widetext}
It is then clear that (\ref{UpsilonABCD}) can be written as:

\begin{equation}
 \upsilon_{ab,cd} = \sum_{I=1}^N  \upsilon_{ab,cd}^I + \sum_{I\neq J}^N \Gamma^I_{ab} \Gamma^J_{cd}
\end{equation}

Similar calculations give the following results:

\begin{eqnarray}
\upsilon_{a,b}&=& - \upsilon_{ab} = \Gamma_{ab}= \sum_{I=1}^N \Gamma^I_{a\,b}\label{vab} \\
\upsilon_{ab\,,\,c}  &=& \sum_{I=1}^N \upsilon_{ab\,,\,c}^I\\
\upsilon_{abc\,,\,d}  &=& \sum_{I=1}^N \upsilon_{abc\,,\,d}^I\\
\upsilon_{abc} &=& \sum_{I=1}^N \upsilon_{abc}^I
\end{eqnarray}
\begin{eqnarray}
\upsilon_{ab\,,\,cd}&=& \sum_{I=1}^N \upsilon_{ab\,,\,cd}^I + \sum_{I\neq J} \Gamma_{a\,b}^I \Gamma_{c\,d}^J\\
\upsilon_{abcd}&=& \sum_{I=1}^N\upsilon_{abcd}^I\\
\upsilon_{ab\,,\,c\,,\,d}&=& \sum_{I=1}^N \upsilon_{ab\,,\,c\,,\,d}^I - \sum_{I\neq J} \Gamma_{a\,b}^I \Gamma_{c\,d}^J\\
\\
\upsilon_{abc\,,\,de} &=& \sum_{I=1}^N \upsilon_{abc\,,\,de}^I - \sum_{I\neq J} \Gamma^I_{de}v^J_{abc}\\
\upsilon_{abcd\,,\,e} &=& \sum_{I=1}^N \upsilon_{abcd\,,\,e}^I\\
\upsilon_{abc\,,\,d\,,\,e} &=& \sum_{I=1}^N \upsilon_{abc\,,\,d\,,\,e}^I + \sum_{I\neq J} \upsilon^I_{abc}\Gamma^J_{de}\\
\upsilon_{ab\,,\,cd\,,\,e}&=&  \sum_{I=1}^N \upsilon_{ab\,,\,cd\,,\,e}^I -\sum_{I\neq J} \left(\upsilon^I_{ab,e}\Gamma^J_{cd}+\upsilon^I_{cd,e}\Gamma^J_{ab}\right)\\
\upsilon_{abcde} &=& \sum_{I=1}^N \upsilon_{abcde}^I\label{vabcde}
\end{eqnarray}

where the single-detector $\upsilon^I_{\{\cdot\}}$ have the values given in \cite{Vitale2010}, that we report in appendix \ref{Appendix_upsilon}.
\\

The $\upsilon$ in which cross terms are present can be further simplified. Let us consider again $\upsilon_{ab,cd}$:

\begin{equation}\label{UpsilonABCDbis}
\upsilon_{ab\,,\,cd}= \sum_{I=1}^N \upsilon_{ab\,,\,cd}^I + \sum_{I\neq J} \Gamma_{a\,b}^I \Gamma_{c\,d}^J
\end{equation}

The single detector $\upsilon_{ab,cd}^I$ has the value:

\begin{equation}
 \upsilon_{ab,cd}^I =\langle s_{ab}\,,\,s_{cd}\rangle^I + \Gamma^I_{ab}\Gamma^I_{cd}.
\end{equation}

so that eq. \ref{UpsilonABCDbis} can be expressed as:

\begin{equation}
\upsilon_{ab\,,\,cd}= \sum_{I=1}^N \langle s_{ab}\,,\,s_{cd}\rangle^I + \sum_{I=1}^N \Gamma^I_{ab}\Gamma^I_{cd}+ \sum_{I\neq J} \Gamma_{a\,b}^I \Gamma_{c\,d}^J
\end{equation}

It is nearly evident that:

$$\sum_{I=1}^N \Gamma^I_{ab}\Gamma^I_{cd}+ \sum_{I\neq J} \Gamma_{a\,b}^I \Gamma_{c\,d}^J = \Gamma_{ab}\Gamma_{cd}\;,$$

which implies that only the network Fisher information appears in the final form:

\begin{equation}
\upsilon_{ab\,,\,cd}= \sum_{I=1}^N \langle s_{ab}\,,\,s_{cd}\rangle^I + \Gamma_{ab}\Gamma_{cd}
\end{equation}

The same kind of calculations show that no cross-terms are present in any of the $\upsilon$, whose final form is:

\begin{eqnarray}
\upsilon_{a,b}&=& - \upsilon_{ab} = \Gamma_{ab}= \sum_{I=1}^N \Gamma^I_{a\,b}\label{vab-final} \\
\upsilon_{ab\,,\,c}  &=& \sum_{I=1}^N \upsilon_{ab\,,\,c}^I\\
\upsilon_{abc\,,\,d}  &=& \sum_{I=1}^N \upsilon_{abc\,,\,d}^I\\
\upsilon_{abc} &=& \sum_{I=1}^N \upsilon_{abc}^I
\end{eqnarray}
\begin{eqnarray}
\upsilon_{ab\,,\,cd}&=&  \sum_{I=1}^N \langle s_{ab}\,,\,s_{cd}\rangle^I + \Gamma_{ab}\Gamma_{cd}\\
\upsilon_{abcd}&=& \sum_{I=1}^N\upsilon_{abcd}^I\\
\upsilon_{ab\,,\,c\,,\,d}&=& - \Gamma_{a\,b} \Gamma_{c\,d}\\
\upsilon_{abc\,,\,de} &=& \sum_{I=1}^N \langle s_{abc} \,,\,s_{de}\rangle^I  - v_{abc} \Gamma_{de}\\
\upsilon_{abcd\,,\,e} &=& \sum_{I=1}^N \upsilon_{abcd\,,\,e}^I\\
\upsilon_{abc\,,\,d\,,\,e} &=& v_{abc} \Gamma_{de} \\
\upsilon_{ab\,,\,cd\,,\,e}&=& -\Gamma_{ab} v_{cd\,,\,e} - \Gamma_{cd} v_{ab\,,\,e}\\
\upsilon_{abcde} &=& \sum_{I=1}^N \upsilon_{abcde}^I.\label{vabcde-final}
\end{eqnarray}

The fact that we have been able to prove that the $\upsilon$'s  explicitly  contain the network Fisher matrix is important, as it allows to obtain simplified expressions for the first order variance and bias:

\begin{equation}\label{SigmaOne}
\sigma^2_{\vartheta^r}[1]=\Gamma^{rr}
\end{equation}

\begin{eqnarray}\label{IMR-BiasOne}
b_{\vartheta^r}[1]&=&\frac{1}{2}\Gamma^{ra}\Gamma^{bc}(\upsilon_{abc}+2\upsilon_{c,ab})
\end{eqnarray}

the second order variance:
\bwt
\begin{eqnarray}\label{IMR-VarMatrixSimplified}
\sigma^2_{\vartheta^j}[2] &=& \Gamma^{jm}\Gamma^{jn}\Gamma^{pq}(\upsilon_{nmpq}\!+\!3 \sum_{I=1}^N \langle s_{nq}\,,\,s_{pm}\rangle^I + 2\upsilon_{nmp,q}\!+\!
\upsilon_{mpq,n}) +\\
&+& \Gamma^{jm}\Gamma^{jn}\Gamma^{pz}\Gamma^{qt}\bigg( v_{npm} v_{qzt} + \frac{5}{2} v_{npq} v_{mzt} + 2 v_{qz\,,\,n}v_{mtp} +2 v_{qp,z} v_{nmt} +\nonumber \\
&+&  6 v_{mqp}v_{nt\,,\,z} + v_{pqz} v_{nt\,,\,m} + 2 v_{mq\,,\,z} v_{pt\,,\,n} +2 v_{pt\,,\,z}v_{mq\,,\,n} + v_{mz\,,\,t}v_{nq\,,\,p}\bigg)\nonumber 
\end{eqnarray}
\ewt

and the second order bias and covariance matrix, that we show in appendix \ref{Appendix_bias2}.
\section{Results}
\subsection{Gravitational waveforms}\label{Section_WF}

In this paper we analyse the errors in the estimation of the physical parameters of a GW signals generated by the inspiral phase of binary systems whose components are not rapidly spinning. We assume that when the frequency of signal enters in the bandwidth of the advanced detectors, these systems have already lost their orbit eccentricity. In this case the plus and cross polarizations are (\cite{Thorne300}, \cite{Schutz2009}):

\begin{eqnarray}
 h(t)_+ &=& A(t) \frac{1+\cos^2\epsilon}{2} \cos \Phi(t) \label{Plus_PN}\\
h(t)_\times &=& A(t) \cos\epsilon \sin \Phi(t) \label{Cross_PN}
\end{eqnarray}

where $A(t)$ is a time dependent amplitude, $\Phi(t)$ is the phase. Both these quantities can be calculated with a higher degree of precision within the frame of the Post-Newtonian theory (\cite{Blanchet2006}, \cite{Damour2005}). The angle  $\epsilon$ is defined later in the text.
The signal (\ref{signal_final}) can be, in this case, written in the form:

\begin{equation}\label{Signal_ThreeDetectors}
 s^I(t)= \mu^I\, A(t-\tau_I) \cos(\Phi(t-\tau_I) - \Phi_0^I)
\end{equation}
where :

\begin{eqnarray}
 \mu^I &\equiv& \sqrt{\left(\frac{1+\cos^2\epsilon}{2} \mathcal{F}_+^I\right)^2  + (\cos\epsilon\, \mathcal{F}_\times^I)^2}\label{rho}\\
\Phi_0^I&\equiv&\ \arctan{ \frac{2 \cos \epsilon\, \mathcal{F}_\times^I}{(1+\cos^2\epsilon )\mathcal{F}_+^I}}\label{varphi_0}
\end{eqnarray}

with $\mathcal{F}_+^I$ and $\mathcal{F}_\times^I$ defined in eqs. (\ref{Generalized_antenna_p}) and (\ref{Generalized_antenna_c}).
The Fourier transform of the signal (\ref{Signal_ThreeDetectors}), considering only positive frequencies, is:

\begin{eqnarray}
  s^I(f) &\propto& \int_{-\infty}^{\infty}{dt e^{- 2\pi i t f} A(t-\tau_I) \left(e^{-i (\Phi(t-\tau_I) -\Phi_0^I)}+e^{i (\Phi(t-\tau_I) -\Phi_0^I)}\right)} \nonumber\\
&=& \int_{-\infty}^{\infty}{dt A(t-\tau_I) e^{- 2\pi\,i t f +i \Phi(t-\tau_I) - i \Phi_0^I}}= \nonumber\\
&=&  e^{-2\pi i f \tau_I - i \Phi_0^I}\,\int_{-\infty}^{\infty}{dT A(T) e^{-2\pi\,i T f +i \Phi(T)}}
\end{eqnarray}

The integral can be solved using the well known stationary phase approximation - SPA -  (\cite{Arun2005}, \cite{Damour2005}) which consists in expanding the integrand around its stationary point, where the derivative of the phase is zero. This also explains why we only kept $e^{-i (\Phi(t-\tau_I) -\Phi_0^I)}$ while going from the first to the second line: the second term would results in an second integral whose integrand would have a phase:

\begin{equation}
\left(-2\pi f t -\Phi(t-\tau) + \Phi_0\right)
\end{equation}

But the derivative of this expression with respect to the time is never zero, because the derivative of the orbital phase $\Phi(t)$ is always positive. This implies that the phase of this additional addend would be oscillating in the whole range of integration, making the integral small.  The final result is then

\begin{equation}
 s^I(f)= \frac{M^\frac{5}{6}}{\pi^\frac{2}{3} D} \sqrt{\frac{5\eta}{24}}\,\mu\,  f^{-\frac{7}{6}} \,e^{i\,\psi(f) - 2\,\pi\,i\, f \tau_I - i\,\Phi_0^I}
\end{equation}

where the phase is given at the 3.5 PN order by:

\begin{eqnarray}
 \psi(f)&=& 2\pi f t_c + \phi_c - \frac{\pi}{4} + \frac{3}{128 \eta v^5}  \sum_{k=0}^7 \alpha_k v^k
\end{eqnarray}

and $v= (\pi M f)^{\frac{1}{3}}$. The coefficients $\alpha_i$, that depend on the total and symmetrized mass, can be found in \cite{Arun2005} and \cite{ArunCorrection}.

Let us summarize the unknown parameters on which this waveform depends:

\begin{itemize}
\item $t_c$ and $\phi_c$ are a reference time (usually the detection time, or the coalescence time) and the phase the wave had at that time
\item $M=m_1 +m_2$ is the total mass, $\eta =\frac{m_1 m_2}{(m_1 +m_2)^2}$ is the symmetric mass ratio. The \emph{chirp mass} $\mathcal{M} = \eta^\frac{3}{5} M$ is often used instead of the total mass.
\item D is the luminosity distance of the system.
\item $\psi$ is the polarization angle
\item $\epsilon$ is the angle formed between the line of sight and the system orbital angular momentum (often referred as iota in the literature). A system with $\epsilon=0$ is called face-on; one with $\epsilon=\pi/2$ edge-on. Note that, because of eq. (\ref{Cross_PN}), an edge-on system only has the plus polarization.
\item $\Theta$ and $\Phi$ are the spherical coordinates of the sources in the fide frame.
It is worth recalling that  $\tau_I$ , $\mu$ and $\Phi_0$ are functions of those angles angles through (\ref{time_shift}),(\ref{Generalized_antenna_p}) ,(\ref{Generalized_antenna_c}),(\ref{rho}) and (\ref{varphi_0}).
\end{itemize}
This waveform is usually assumed to be accurate till the \emph{innermost stable circular orbit} (ISCO) frequency. This will also be the upper limit of integration in the noise-weighted integrals like (\ref{Fourier_Mean}):

\begin{equation}
f_{ISCO} = f_{up} = \left(6^\frac{3}{2} \pi M\right)^{-1}
\end{equation}

It is know, and easy to verify, that the distance and $\epsilon$ are strongly correlated. From a numerical point of view this makes the inversion of the Fisher information matrix (\ref{FisherNetw}) numerically unstable, the determinant being close to zero. This is a known problem (\cite{Vallisneri2008}) with explicit or emerging degeneracies. 
Numerical issues need to be handled carefully since the second order of the covariance expansion
(\ref{Covariance}) contain the IFIM multiplied by itself up to four times. 
To prevent numerical instabilities, in this paper we consider two parameters spaces: 
one where only the position of the source is unknown (mostly to compare our results with existing literature
that adopt this simplification), and  $\bm{\vartheta}=(t_a,log\mathcal{M},\eta,lat,long)$. 
We checked that in order to extend the parameter space further (for example up to 9 parameters)
the numerical precision would need to exceed one hundred digits.

\subsection{Detectors locations and noise models}

In table \ref{TablePosition} we report the position of the detectors beam splitter \cite{Schutz2011} with respect to the Earth Frame (eq. (\ref{EF_to_DF}) can be used to convert those in the Euler angles needed to calculate the generalized antenna patterns).
More details on the existing and planned IFOs can be found in (\cite{ILigo,ALigo},\cite{IVirgo,IVirgo2,AVirgo}, \cite{AIGO}, \cite{Indigo}, \cite{LCGT}).

\begin{table}[h!b!p!]
\begin{tabular}{|c|c|c|c|c|}
\hline
 Detector & Label & Longitude & Latitude & $\Delta$\\
 \hline
LIGO Livingston & L& $90^\circ 46' 27.3''$ W & $30^\circ 33' 46.4''$ N & $108^\circ0'0''$ \\
LIGO Hanford & H& $119^\circ 24' 27.6''$ W & $46^\circ 27' 18.5''$ N& $36.8^\circ0'0''$\\
Virgo, Italy &V& $10^\circ 30' 16''$E & $43^\circ 37'53''$ N & $341^\circ30'0''$\\
LCGT, Japan & J& $137^\circ 10' 48''$ E & $36^\circ15'00''$ N & $295^\circ 0'0''$\\
AIGO,  Australia & A & $115^\circ42'51''$E& $31^\circ21'29''$ S & $270^\circ0'0''$\\
INDIGO, India & I & $74^\circ02'59''$E& $19^\circ05'47''$ S & $45^\circ0'0''$\\
\hline
\end{tabular}
\caption{ The positions and orientations of the detector in the Earth Frame. The angle $\Delta$ is defined as the angle between the local North and the first arm of the interferometer measured counter clockwise. Eq. (\ref{Euler_to_Latitude}) can be used to convert those angles in the Euler angles that transform the Detector Frame in the Earth Frame.}\label{TablePosition}
\end{table}

\begin{figure}[htb]
\includegraphics[scale=\scale]{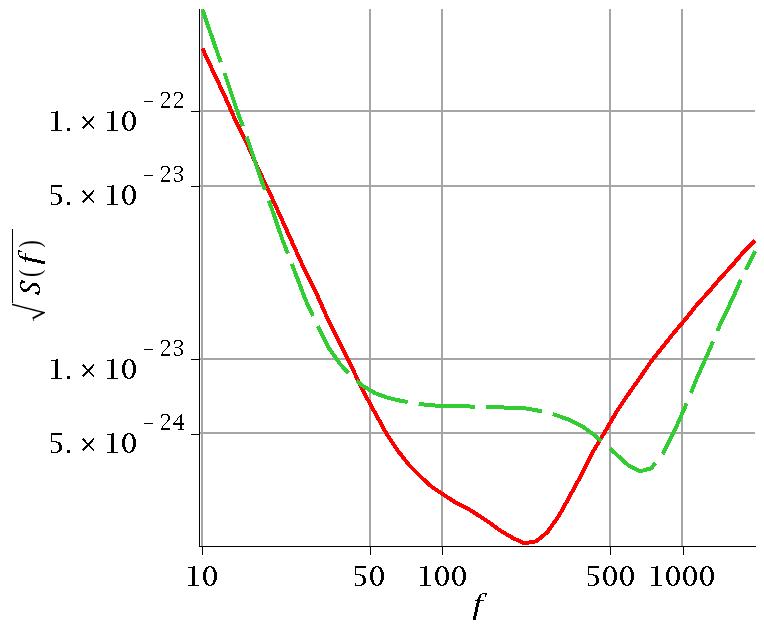} 
\caption{(Color online) Square root of the noise PSD for an Advanced Ligo detector (continuous line) an Advanced Virgo detector (dotted line)}\label{Fig_noises}
\end{figure}

For the advanced Virgo detector we have used the noise power spectral density plotted in Fig. \ref{Fig_noises} and given in \cite{Vitale2010}:
\bwt
\begin{eqnarray}\label{AdvVirgo}
 S_h(f)&=& S_0 \left[ 2.67\,10^{-7}\, x^{-5.6} +0.59\, e^{(\ln{x})^2\,\left[-3.2 -1.08 \ln{x} -0.13 (\ln{x})^2\right]} x^{-4.1} +0.68\, e^{-0.73\, (\ln{x})^2 } x^{5.34}\right],\; f\geq f_{low}\nonumber\\
 S_h(f)&=&\infty, \;\;f\leq f_{low}
 \end{eqnarray}
\ewt
 Where $x\equiv \frac{f}{f_0}$, $f_0= 720 \mbox{Hz}$, and $S_0= 10^{-47} \mbox{Hz}^{-1}$. 
 
 or the advanced LIGO detector, as well as for the projected new IFOs, we use the following noise power spectral density, also plotted in Fig. \ref{Fig_noises}:
 
\begin{eqnarray}\label{AdvLigo}
 S_h(f)&=& S_0 \left[x^{-4.14} -5 x^{-2} + 111\frac{1-x^2+x^4/2}{1+x^2/2}\right],\; f\geq f_{low}\nonumber\\
 S_h(f)&=&\infty, \;\;f\leq f_{low}
 \end{eqnarray}
Where $x\equiv \frac{f}{f_0}$, $f_0= 215 \mbox{Hz}$, and $S_0= 10^{-49} \mbox{Hz}^{-1}$. 

For both the noise spectral densities, we have chosen the lower frequency cutoff to be $f_{low}=20\mbox{Hz}$.

Obviously the noise for the new detectors need not to be exactly like this one, but using the AdvLIGO configuration is a good approximation for the sensitivities of those new detectors, and the differences will be very small compared to the addition of a new IFO to the network.

\subsection{Representative examples and figures of merit}
In the frequentist formalism the errors depend on the value of the parameters. 
Therefore it is informative to show which kind of errors, 
calculated using the first two terms of eq. (\ref{Sigma_series}),
we can expect for different representative signals and different locations in the sky.
It is also important to keep in mind that some situations are intrinsically 
pathological and less interesting: for example if the latitude is 
0 or $\pi$ the longitude is undefined so it is not 
meaningful to compute the errors on it (the FIM would actually be singular). 
In this paper we do not quote results for the bias, eq. (\ref{bias_series}), 
as we shown in \cite{Vitale2010} that those are important only for systems 
much more massive than the ones we consider in this work.

The plots in this paper refer to the following binary systems:
\begin{itemize}
\item BNS - Binary neutron star - $m_1=m_2=1.4 M_{\odot}$
\item BHNS - Black hole - neutron star - $m_1=10 M_{\odot}, m_2=1.4 M_{\odot}$
\item BBH - Binary black hole - $m_1=m_2=10 M_{\odot}$
\end{itemize}
and network configurations:
\begin{itemize}
\item 3 Detectors: HLV 
\item 4 Detectors: HLVJ, HLVA and HLVI
\item 5 Detectors: HLVJA and HLVJI
\end{itemize}

When we calculate the errors, unlike some existing studies (e.g. \cite{Fairhust2010}), we do not fix the distance of the sources to be the same for all the networks. The reason is that such approach does not disentangle the role of SNR gains and network geometry (see some related discussion in \cite{Aylott2011}).
Being more interested in the role of the network geometry, for each network 
and mass bin we set the distance of the source so that the average network SNR over the sky
is about 15. In particular: 

\begin{itemize}
\item 3 Detectors: BNS: 200Mpc, BHNS: 450Mpc, BBH:1100Mpc
\item 4 Detectors: BNS: 250Mpc, BHNS: 550Mpc, BBH:1300Mpc
\item 5 Detectors: BNS: 280Mpc, BHNS: 6250Mpc, BBH:1500Mpc
\end{itemize}

Note however how the the joint effect of additional IFOs and higher SNR can be quantified from the results we quote in this paper. As we provide both $\sigma[1]$ and $\sigma[2]$, as well as the network SNR, $\rho^*$, for which they are calculated, one can build for any SNR, $\rho$, the quantity (which is easily derived from eq. \ref{Sigma_series}):
\beq
\sigma^2(\rho) = \sigma^2[1] \left(\frac{\rho^*}{\rho}\right)^2 + \sigma^2[2] \left(\frac{\rho^*}{\rho}\right)^4
\eeq

which gives the errors for the desired SNR. 

In order to study the sky dependence of the angular resolution 
we adopt a 289 points sky grid where the $(i,j)$-th point has latitude and longitude
in the Earth frame.
\begin{eqnarray}
lat_{ij}&=& \left[\frac{\pi}{2} -\delta \right] - \frac{\pi +2 \delta}{16} j \nonumber\\
long_{ij}&=& -\pi+\delta  + \frac{2\pi - 2\delta}{16} i\nonumber\\ \nonumber	 \\
\mbox{with } i,j &=& 0,1,\cdots ,16 \;;\; \delta=0.02.
\end{eqnarray}

In each point the Fisher matrix and the second order corrections are evaluated.
The small offset $\delta$ in the range of the latitude is added to avoid the poles.
The first and last values of the latitude are then $\frac{\pi}{2} -\delta$ and $-\frac{\pi}{2} +\delta$.

In our simulation we set to representative values of the orientation angle $\epsilon= \pi/6$ and $\psi=0$. 
Even if the results are almost independent of $\psi$
the choice of the values of $\epsilon$ deserve some discussion.
It can be shown, from the derivatives of \ref{rho} and \ref{varphi_0} with respect to $\epsilon$,
that for $\epsilon=0$ derivative of the signal with respect to $\epsilon$ is zero
and therefore the Fisher matrix is singular. Similar inversion problems are present for small but non zero values
of $\epsilon$. However, apart from the neighbourhood of zero, the results would not be qualitatively different choosing different values of $\epsilon$.  We explicitly varied $\epsilon$, keeping the other parameters fixed, and observed 
only changes in SNR due to the fact that the cross polarization become less and less important (\ref{Cross_PN}). 
By rescaling the distance to compensate for this loss, the pattern on the errors in the sky-map are roughly unchanged. 
This is also the case with the larger networks we considered, where the polarization angle can be partially resolved, and the pattern on the errors in the sky-map is pretty much unchanged.

In the following we describe how the accuracies in the estimation of the direction of arrival are be presented.
The covariance matrix and its second order give us the errors in latitude, longitude, and their correlation.
Assuming that the likelihood is roughly ellipsoidal the probability distribution of the angular position is a bivariate Gaussian that can be written as:
\bwt
\begin{equation}
p(\Delta lat,\Delta long)\\ \nonumber
 \propto e^{-\frac{1}{2} \left( \Delta lat^2 \Gamma_{lat,lat} + \Delta long^2 \Gamma_{long,long}
 +  2 \Delta lat \Delta long \Gamma_{lat,long}\right)}
\end{equation}
\ewt
Where $\Gamma$ is the inverse of the covariance matrix (or FIM if only the first order is being taken into account).
The contours of constant probability are ellipses in the (long,lat) plane. Because of the correlation term, the axis of the ellipse are not parallel to the coordinate axes but form an angle $\alpha$.

One can calculate the principal directions of the ellipses diagonalizing the matrix:

\begin{equation}
\left( \begin{matrix}
\Gamma_{long,long} &\Gamma_{lat,long} \\
\Gamma_{lat,long}&\Gamma_{lat,lat} 
\end{matrix}\right)
\end{equation}

and finding eigenvalues $\lambda_1$ and $\lambda_2$. It is evident that the widths of the ellipse along the two eigenvectors are then simply $ \delta_i = \frac{1}{\sqrt{\lambda_i}}$, with $i=1,2$. The solid angle corresponding to one of such ellipses, centered around a point of latitude $\gamma$ can be written as:
\begin{equation}\label{Solid_Angle_error}
\Delta\Omega= 2\pi |\cos \gamma| \frac{1}{\sqrt{\lambda_1 \lambda_2}}
\end{equation}
Considering the scalar product of the first eigenvector with the unit vector of the $long$ axis one can find the angle $\alpha$. These ellipses are then plotted using an Aitoff projection of the sky.

\subsection{Bidimensional Parameter Space}

In this section we compute the uncertainties for the location of a binary system, while the other parameters are assumed known.  The approximation of reducing the parameter space, used for example in \cite{Fairhust2009},\cite{Fairhust2010}, \cite{Wen2010}, 
unfortunately tends to produce optimistic predictions as discussed in details in the next section. 
The performance of the different network configurations considered are summarized in tables \ref{Table2Dim3Det}, \ref{Table2Dim4Det} and \ref{Table2Dim5Det}. 

While the entries that use just the IFIM in table II and 
for the HLVA network in table III are consistent with the results in 
\cite{Wen2010}, \cite{Fairhust2010}, columns II IV and VI show that in realistic
SNR regimes the second order needs to be taken into account.  
For example, if we include the second order for an average network SNR of 15, 
the solid angle uncertainty increases on average $45\%$, or more in detail
between $37\%$ and $137\%$ with 3 IFOs,
between $10\%$ and $164\%$ with 4 IFOs and
between $10\%$ and $153\%$ with 5 IFOs .

These tables indicate the benefits of enlarging the IFO 
network however we defer interpretation and discussions to the next section where 
intrinsic parameters are also estimated at the same time as the direction of arrival.

\begin{table}[h!b!p!]
\begin{tabular}{|c|c|c|c||c|c||c|c|}
\hline
&&\multicolumn{2}{c||}{BNS}&\multicolumn{2}{c||}{BHNS}&\multicolumn{2}{c|}{BBH}\\
&& $\sigma[1]$ & $\sigma[2]$ & $\sigma[1]$ & $\sigma[2]$& $\sigma[1]$ & $\sigma[2]$\\
 \hline
\multirow{3}{3mm}{\begin{sideways}\parbox{15mm}{HLV}\end{sideways}}&lat [mrad]    & 7.69 & 6.18    &11.3 & 6.01 &15.3 & 6.85\\  \cline{2-8}
&long [mrad]              & 15.1 & 8.13   &20.3 &87.92   &25.6 & 8.86\\ \cline{2-8}
&$\Delta\Omega [deg^2] $ & 0.84 & 2.23*     & 1.55 &2.99* & 2.80&4.45*\\  \cline{2-8}
&Net. SNR & \multicolumn{2}{c||}{16.55} & \multicolumn{2}{c||}{15.37} & \multicolumn{2}{c|}{14.26}\\
\hline
\end{tabular}
\caption{ First order, $\sigma[1]$, and second order, $\sigma[2]$, errors averaged over the sky using a three detectors network. Only the sky position is considered unknown \\ * Note: for the solid angle error, the column $\sigma[2]$ takes into account both the first and second order errors. The reason is that for the solid angle error, eq. (\ref{Solid_Angle_error}), is not possible to separate the effects of first and second order.}\label{Table2Dim3Det}
\end{table}

\begin{table}[h]
\begin{tabular}{|c|c|c|c||c|c||c|c|}
\hline
&&\multicolumn{2}{c||}{BNS}&\multicolumn{2}{c||}{BHNS}&\multicolumn{2}{c|}{BBH}\\
&& $\sigma[1]$ & $\sigma[2]$ & $\sigma[1]$ & $\sigma[2]$& $\sigma[1]$ & $\sigma[2]$\\
 \hline
\multirow{3}{3mm}{\begin{sideways}\parbox{15mm}{HLVA}\end{sideways}}&lat [mrad]& 7.96 & 6.09  &10.9 & 5.02 &14.3 & 5.53\\ \cline{2-8}
&long [mrad]              & 14.2 & 7.31  &18.1  &6.38   &21.9 & 6.82\\ \cline{2-8}
&$\Delta\Omega [deg^2] $ & 0.89  & 1.65*    & 1.47  &2.11* & 2.47&3.19*\\ \cline{2-8}
&Net. SNR & \multicolumn{2}{c||}{15.7} & \multicolumn{2}{c||}{14.94} & \multicolumn{2}{c|}{14.29}\\
\hline\hline
\multirow{3}{3mm}{\begin{sideways}\parbox{15mm}{HLVI}\end{sideways}}&lat [mrad]     & 7.24 & 4.12  &9.33 & 3.30 &12.3 & 3.85\\ \cline{2-8}
&long [mrad]              & 14.7 & 6.16 &17.6 &5.03   &21.0 & 5.65\\ \cline{2-8}
&$\Delta\Omega [deg^2] $ & 0.77 & 1.06*  & 1.17 &1.36*   & 1.99 &2.23*\\ \cline{2-8}
&Net. SNR & \multicolumn{2}{c||}{15.4} & \multicolumn{2}{c||}{14.94} & \multicolumn{2}{c|}{14.05}\\
\hline\hline
\multirow{3}{3mm}{\begin{sideways}\parbox{15mm}{HLVJ}\end{sideways}}&lat [mrad]               & 6.87 & 5.42  &9.09 & 4.56 &11.9 & 5.00\\ \cline{2-8}
&long [mrad]              & 13.5 & 7.39  &16.7 &6.28  &20.4 & 6.82\\ \cline{2-8}
&$\Delta\Omega [deg^2] $ & 0.75   & 2.17*    & 1.18 &2.30*  & 2.02&3.04*\\ \cline{2-8}
&Net. SNR & \multicolumn{2}{c||}{16.01} & \multicolumn{2}{c||}{15.23} & \multicolumn{2}{c|}{14.57}\\
\hline
\end{tabular}
\caption{Same as table \ref{Table2Dim3Det}, but using 4-IFO networks}\label{Table2Dim4Det}
\end{table}

\begin{table}[h]
\begin{tabular}{|c|c|c|c||c|c||c|c|}
\hline
&&\multicolumn{2}{c||}{BNS}&\multicolumn{2}{c||}{BHNS}&\multicolumn{2}{c|}{BBH}\\
&& $\sigma[1]$ & $\sigma[2]$ & $\sigma[1]$ & $\sigma[2]$& $\sigma[1]$ & $\sigma[2]$\\
 \hline
\multirow{3}{3mm}{\begin{sideways}\parbox{15mm}{HLVJI}\end{sideways}}&lat [mrad]             & 6.44  & 3.46 &7.97 & 2.66&10.7  & 3.31\\ \cline{2-8}
&long [mrad]            & 12.9  & 4.86&15.1 &4.08 &18.5 & 4.86\\ \cline{2-8}
&$\Delta\Omega [deg^2]$& 0.65 &  0.85* &0.94  & 1.05*      & 1.67  &1.84*\\ \cline{2-8}
&Net. SNR & \multicolumn{2}{c||}{15.92} & \multicolumn{2}{c||}{14.97} & \multicolumn{2}{c|}{14.06}\\
\hline
\hline
\multirow{3}{3mm}{\begin{sideways}\parbox{15mm}{HLVJA}\end{sideways}}&lat [mrad]              & 6.76 & 4.54 &8.72 & 3.53 &11.7 & 4.24\\\cline{2-8}
&long [mrad]             & 13.1 & 6.04 &16.0  &4.96   &19.7 & 5.80\\ \cline{2-8}
&$\Delta\Omega [deg^2]$ & 0.71  & 1.12*  &1.08     &1.36*    &1.92&2.29*\\ \cline{2-8}
&Net. SNR & \multicolumn{2}{c||}{16.18} & \multicolumn{2}{c||}{15.18} & \multicolumn{2}{c|}{14.28}\\
\hline
\end{tabular}
\caption{Same as table \ref{Table2Dim3Det}, but using 5-IFO networks}\label{Table2Dim5Det}
\end{table}

\subsection{Five dimensional parameter space}

In this section we analyze the errors when a five dimensional parameter space $\bm{\vartheta}=(t_a,log\mathcal{M},\eta,lat,long)$ is used and discuss in details the benefits of enlarging the HLV network of interferometers.
Quantitative results are presented in tables (\ref{Table5Dim3Det}, \ref{Table5Dim4Det} and \ref{Table5Dim5Det}) and 
in figures 2 to 10. The discussion that follow is organized in two lines: (a) trends in the accuracy of the direction reconstruction and  estimation of intrinsic parameters (b) methodology recommendations to obtain reliable estimates. 

\begin{figure}[htb]
\includegraphics[scale=\scale]{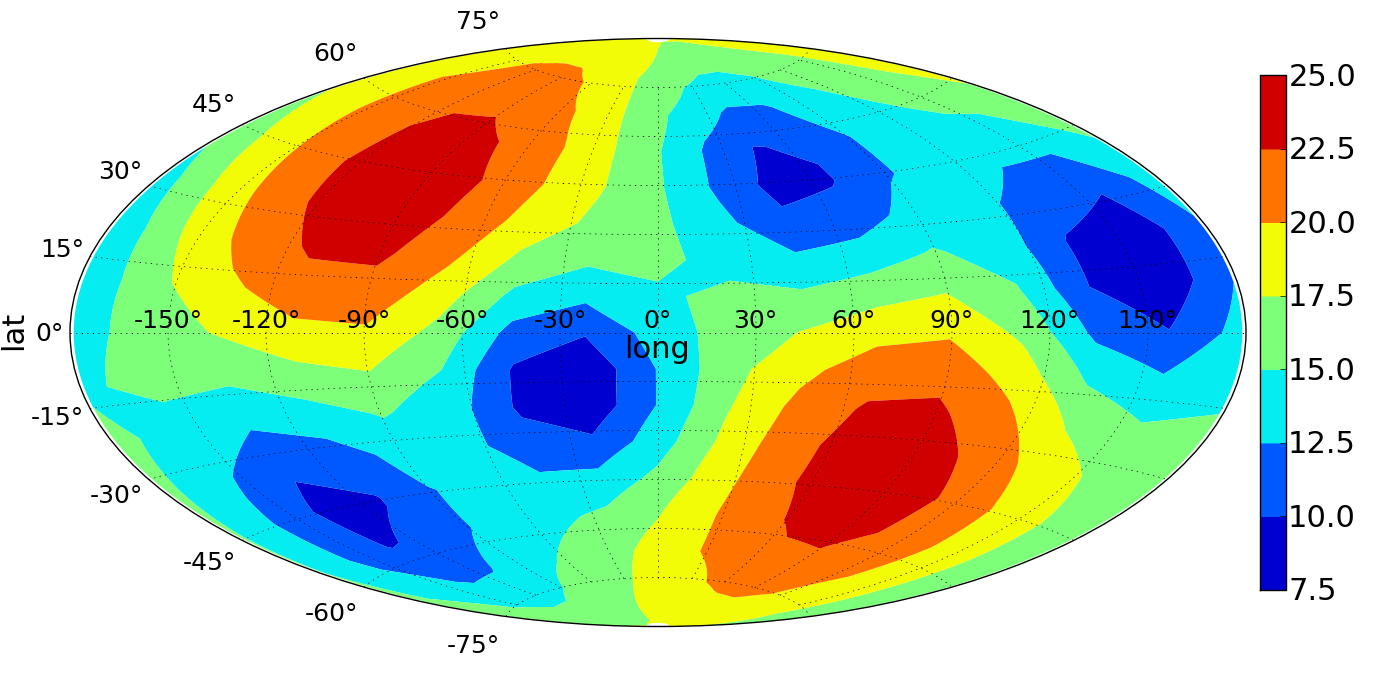}
\caption{(Color online) Network SNR with an HLV network for a BNS system at 200Mpc of distance.}
\label{SNR_HLV_BNS}
\end{figure}

(a)
It is known that with a three detectors network the source sky localization is generally not very good, and that several blind spots exist in the projection on the sky of the plane containing the three IFO. This plane corresponds to the low-SNR region of fig. \ref{SNR_HLV_BNS}.

\begin{figure}[htb]
\includegraphics[scale=\scale]{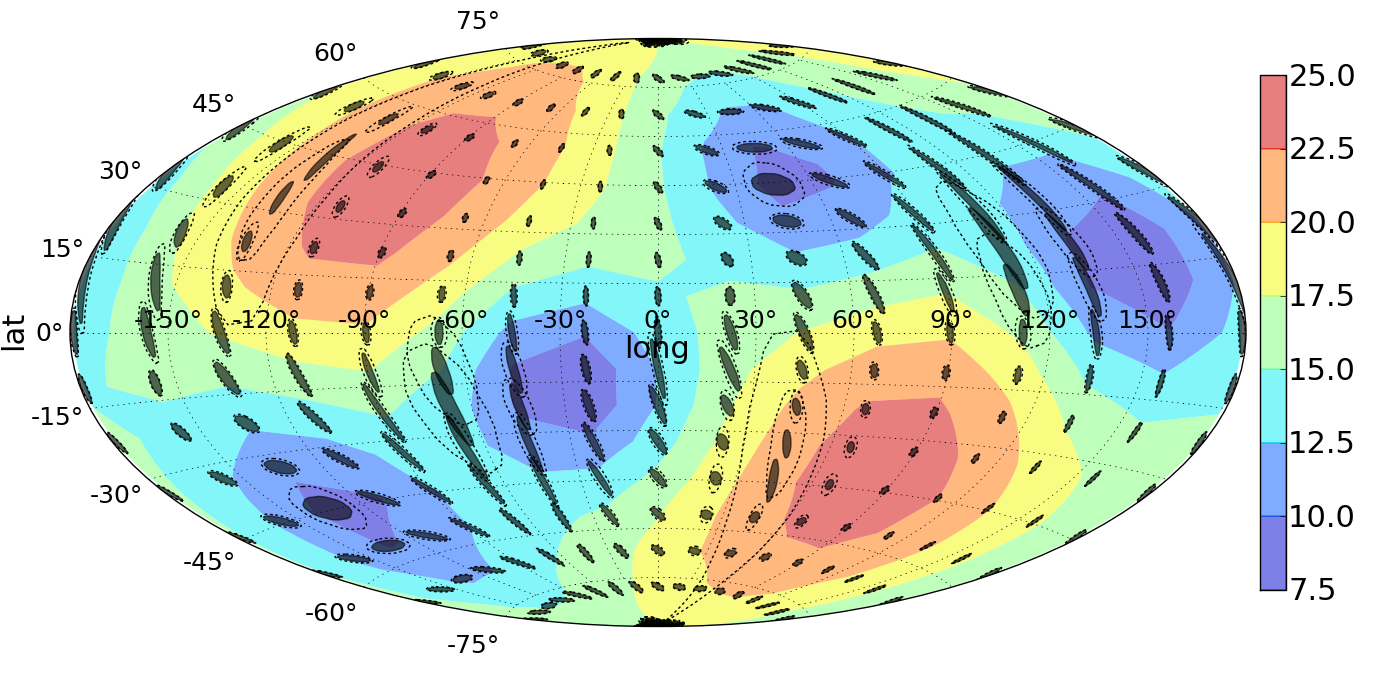}
\caption{(Color online) Skymap of the errors for a BNS signal detected in HLV. The Network SNR is plotted in the background and its value is shown in the colorbar}
\label{SkyHLV_BNS}
\end{figure}

In fig. \ref{SkyHLV_BNS} we plot the ellipses corresponding to 95\% confidence interval for the BNS system; the solid ellipses are obtained using the first order errors, while the dashed ellipses take into account the second order. We have plotted a filled contour of the network SNR in the background as comparison. It is evident how, even though most of the biggest ellipses are in low-SNR regions, as someone might expect, there is no common proportionality between the two, and the contribution of the second order to the sky localization can be big for regions with medium SNR (e.g. long 45 lat 45 S in fig. \ref{SNR_HLV_BNS}). We return to this point later in the section.
In table \ref{Table5Dim3Det} we report the errors, averaged over the whole sky, for the three binary system considered here, when  detected with the HLV network.

\begin{table}[h]
\begin{tabular}{|c|c|c|c||c|c||c|c|}
\hline
&&\multicolumn{2}{c||}{BNS}&\multicolumn{2}{c||}{BHNS}&\multicolumn{2}{c|}{BBH}\\
&& $\sigma[1]$ & $\sigma[2]$ & $\sigma[1]$ & $\sigma[2]$& $\sigma[1]$ & $\sigma[2]$\\
 \hline
\multirow{7}{3mm}{\begin{sideways}\parbox{15mm}{HLV}\end{sideways}}& $\mathcal{M}$ [\%]   & 4.31e-3& 2.58e-3 & 2.66e-2 & 1.68e-2 &0.21 & 0.14\\ \cline{2-8}
&$\eta$ [\%]            & 0.22 & 0.14      & 0.38 & 0.21 & 1.61  & 1.00 \\ \cline{2-8}
&$t_a$ [ms]              & 0.34 & 0.23     & 0.46& 0.27 & 0.68& 0.39\\ \cline{2-8}
&lat [mrad]               & 22.5 & 17.9    &28.6 & 19.7 &36.1 & 22.0\\ \cline{2-8}
&long [mrad]              & 51.9 & 46.1    &57.0 &35.3 &76.1 & 48.7\\ \cline{2-8}
&$\Delta\Omega [deg^2] $ & 4.71 & 16.4* & 9.07 &21.5* &15.14 &27.09*\\ \cline{2-8}
&Net. SNR & \multicolumn{2}{c||}{16.55} & \multicolumn{2}{c||}{15.37} & \multicolumn{2}{c|}{14.27}\\
\hline
\end{tabular}
\caption{ First order, $\sigma[1]$, and second order, $\sigma[2]$, errors averaged over the sky using a three detectors network and a five dimensional parameter space\\ * Note: for the solid angle error, the column $\sigma[2]$ takes into accout both the first and second order errors. The reason is that for the solid angle error, eq. (\ref{Solid_Angle_error}), is not possible to separate the effects of first and second order.}\label{Table5Dim3Det}
\end{table}

The sky-average accuracy, at an average network SNR of about 15,
averaged between the three mass bins that we consider here, is 
3.2 square degrees for HLV when only the two dimensional 
parameter space is assumed. However for the same 
network and the five dimensional parameter space this number 
raises to 21.6 square degrees.

Increasing the size of the network to 4 IFOs improve the situation 
substantially with these two numbers becoming
2.1 and 9.0 square degrees (AIGO is the best 
addition with 5.5 square degree for the 5 dimensional parameter space).
Finally if the network includes 5 interferometers 
these uncertainties would decrease to 1.4 and 3.5 square degrees.

It is also important to notice that the angular resolution is not homogeneous 
in the sky (especially with the HLV network). and that the role of the second 
order in the expansions is usually milder in the locations where the accuracy 
is of the order of a degree.  

One way to quantify the fraction of the sky that can provide sufficient directional 
accuracy for electro magnetic follow-ups is described in table \ref{TableSummary}.
For example we can observe how the fraction of the sky (averaged over the 3 
binary signals) where the one sigma angular resolution is below 2 square degrees 
is $20\%$ for the HLV network and increases to $62\%$ for the best performing 
5 IFOs network (HLVJA).

Comparing the results of the four site networks, summarized in table  \ref{Table5Dim4Det}, with the three IFO network, table \ref{Table5Dim3Det}, it is evident that there are not substantial improvements in the estimation of the intrinsic parameters, chirp mass and $\eta$, while the errors in the estimation of the arrival time, and sky position are dramatically lower.

\begin{figure}[htb]
\includegraphics[scale=\scale]{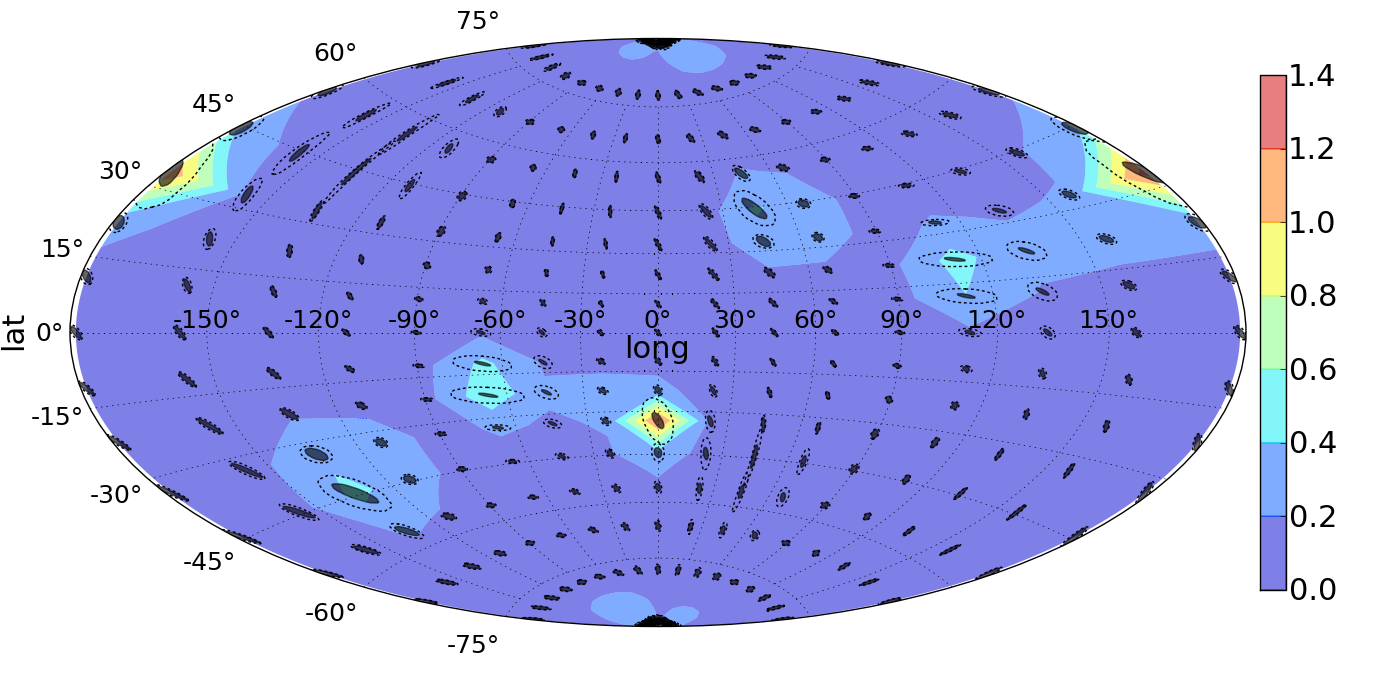}
\caption{(Color online) Skymap for a BNS signal detected in HLVA. The error on the arrival time is plotted in the background and its value is shown in the colorbar (in ms)}
\label{SkyHLVA_BNS}
\end{figure}

Looking at the plots we can also observe that the low sky-error regions do not overlap with high-SNR regions. The reason is that the direction uncertainties mostly depend on the geometry of the network and the estimation of the arrival time. In fig. \ref{SkyHLVA_BNS} we plot the error ellipses corresponding to a BNS signal detected in HLVA, but this time we put the error on the arrival time (first plus second order) as a filled contour background, instead of the network SNR. The match between large errors in time and position is evident. This also implies that if the exact position and orientation of new detectors
is chosen to guarantee the highest network SNR, on average, it might not need not to assure the best performances for sky localization (\cite{Vitale2011}). This fact is confirmed by table \ref{Table5Dim4Det}, from which is clear that HLVJ is the network that guarantee the highest averaged SNR, but not the lowest errors for intrinsic parameters.

\begin{figure}[htb]
\includegraphics[scale=\scale]{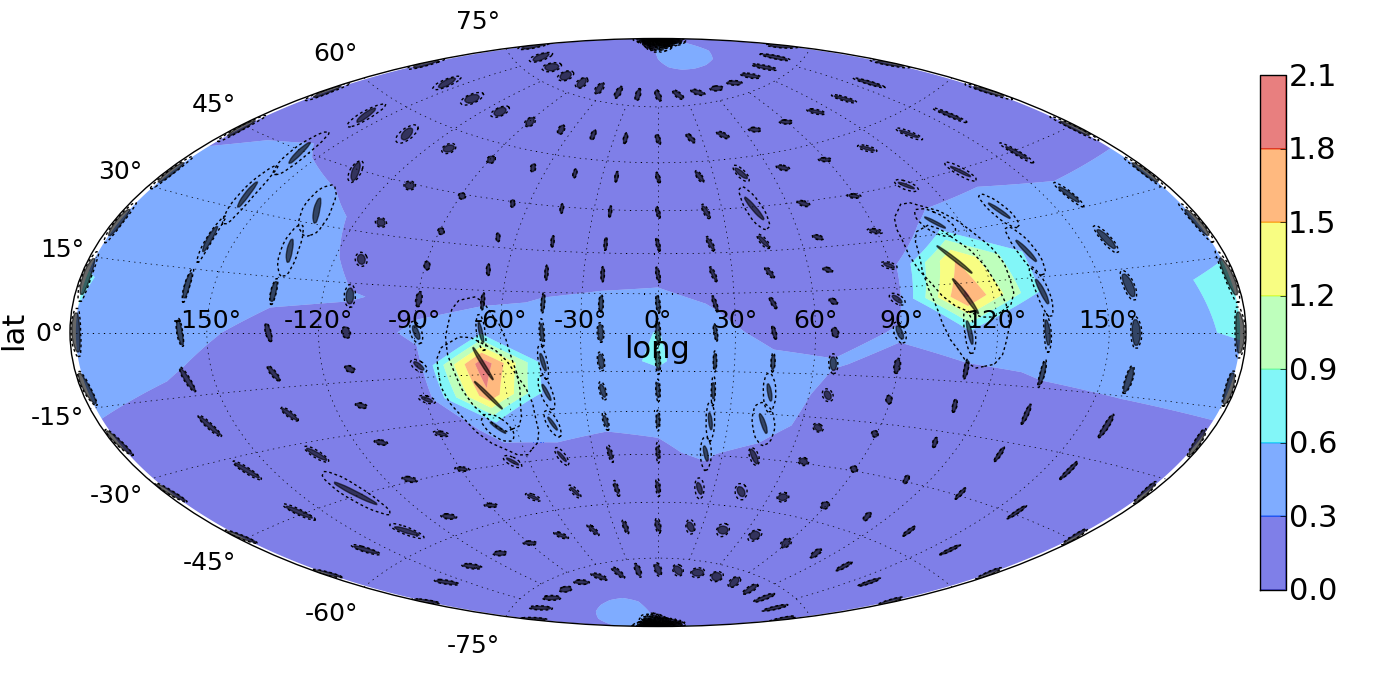}
\caption{(Color online) Skymap for a BNS signal detected in HLVI. The error on the arrival time is plotted in the background and its value is shown in the colorbar (in ms)}
\label{SkyHLVI_BNS}
\end{figure}

\begin{figure}[htb]
\includegraphics[scale=\scale]{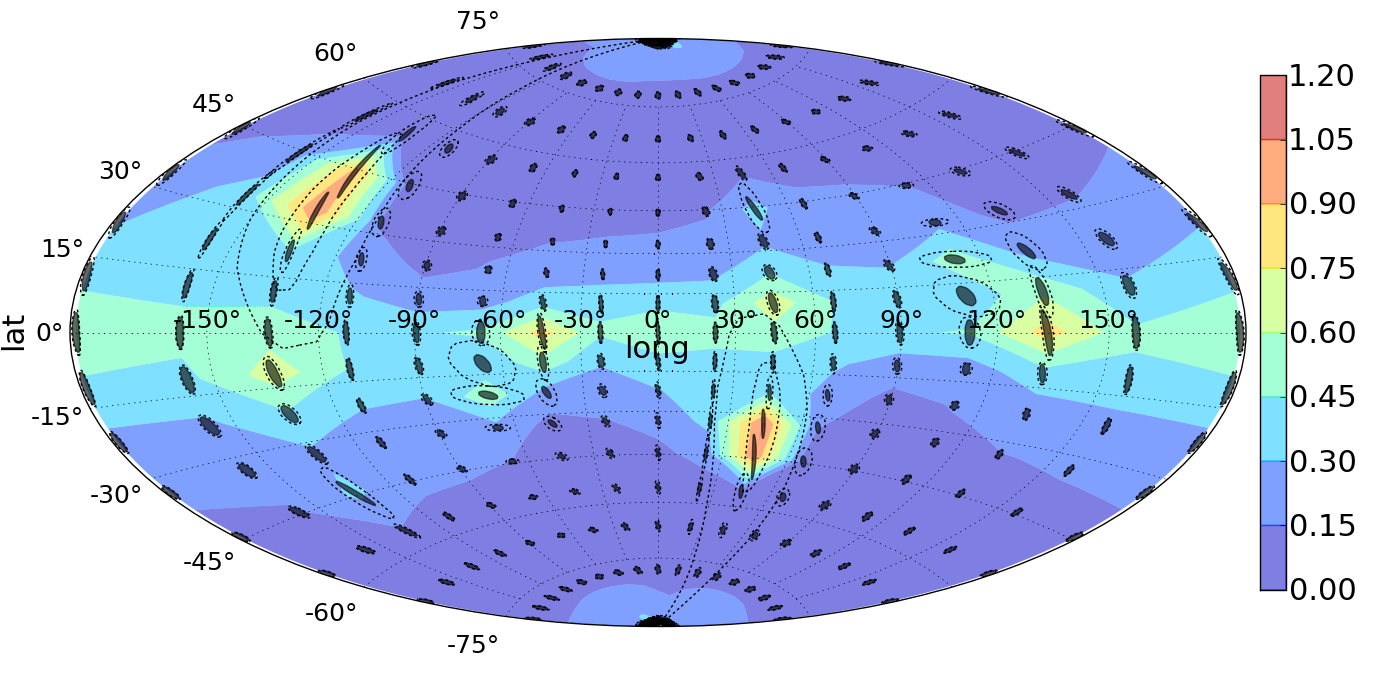}
\caption{(Color online) Skymap for a BNS signal detected in HLVJ. The error on the arrival time is plotted in the background and its value is shown in the colorbar (in ms)}
\label{SkyHLVJ_BNS}
\end{figure}

In fig. \ref{SkyHLVI_BNS} and \ref{SkyHLVJ_BNS} we show errors on the reconstruction of the same source, but using HLVI and HLVJ networks.


\begin{figure}[htb]
\includegraphics[scale=\scale]{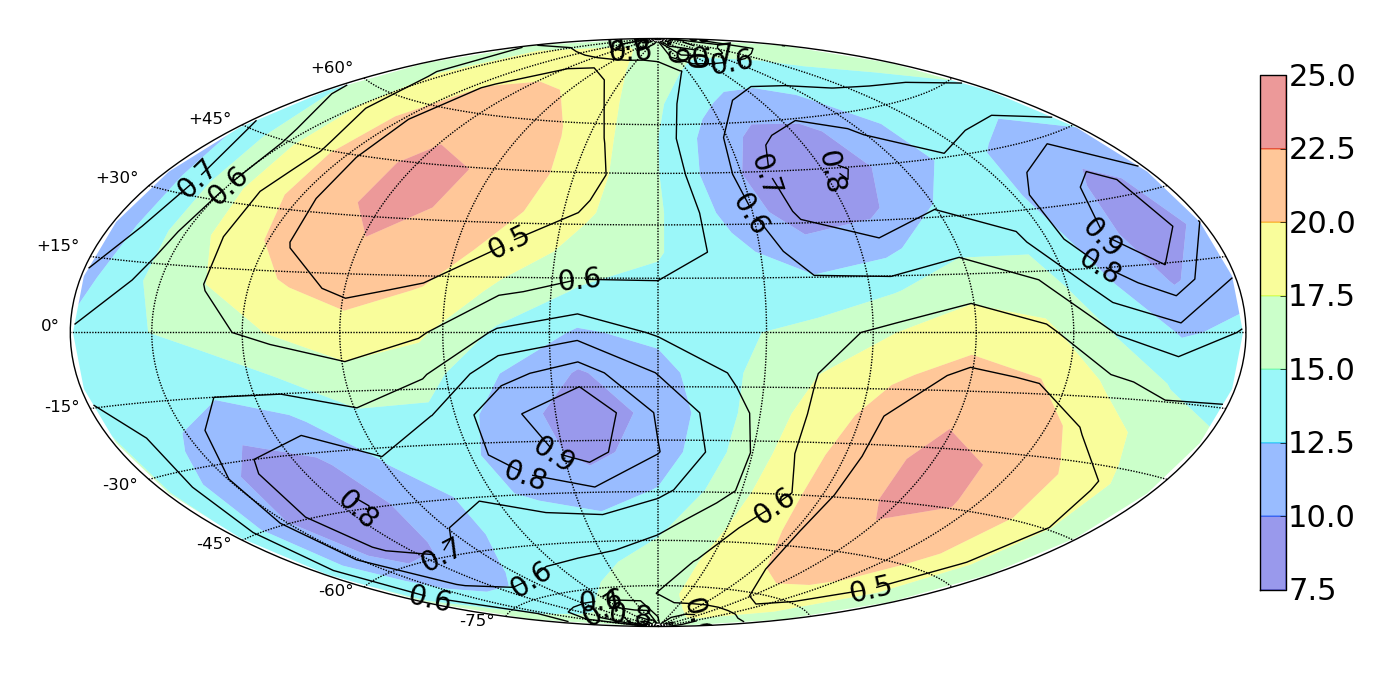}
\caption{(Color online) Ratio between the second and first order errors for the chirp mass. BNS detected in HLVA. Network SNR in background and colormap}
\label{SoverB_logC_HLVA_BNS}
\end{figure}

\begin{figure}[htb]
\includegraphics[scale=\scale]{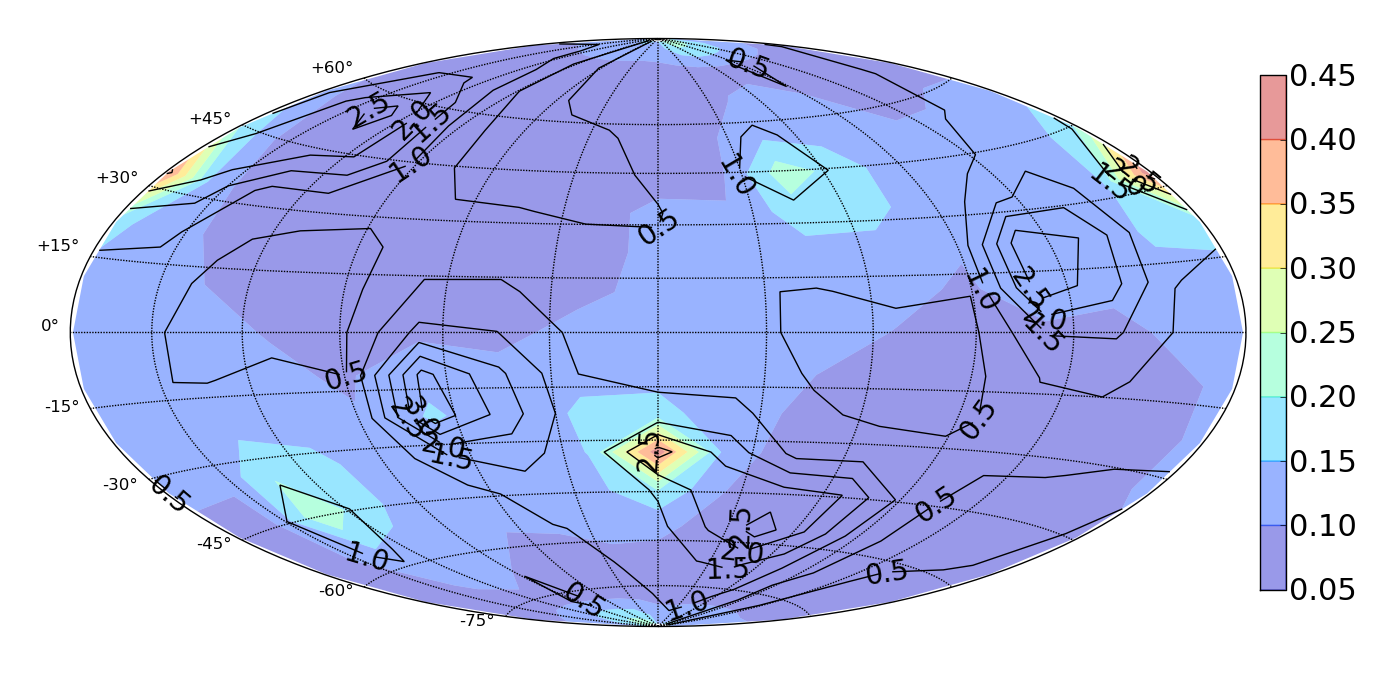}
\caption{(Color online) Ratio between the second and first order errors for the latitude estimation. BNS detected in HLVA. Error in arrival time (ms) in background and colormap}
\label{SoverB_lat_HLVA_BNS}
\end{figure}

Those plots seems to suggest that the HLVA network allows the best sky localization, which is also the outcome of table \ref{Table5Dim4Det}. This is reasonable given that 1) the AIGO site is the furthest from the HLV IFOs, and 2) the plane containing AIGO's arms would be nearly perpendicular to the HLV plane (providing sensitivity in most of the blind spots of the HLV network).
Note that HLVA is also the network for which the second order contributions for $\Delta \Omega$ are smallest. On average the ratio between $\Delta \Omega[1st+2nd]$ and $\Delta \Omega[1st]$ is $3.15$ for BNS, $2.2$ for BHNS and $1.79$ for BBH, using HLVA. The same ratios are respectively $5.17$, $2.76$ and $1.92$ using HLVI and $6.76$, $2.95$ and $2.13$ using HLVJ
For the intrinsic parameters, chirp mass and $\eta$, the second order contributions are larger were the SNR is lower; in fig. \ref{SoverB_logC_HLVA_BNS}, for example, we plot the ratio $\frac{\sigma_\mathcal{M}[2]}{\sigma_\mathcal{M}[1]}$ in contours, with the network SNR in the background and colormap.
For the extrinsic parameters, on the other hand, the second order terms are larger where the error on the first order arrival time were large. This is well shown in fig. \ref{SoverB_lat_HLVA_BNS}, where it is visible that the biggest contributions to the second order errors in latitude come from regions where the arrival time error is large. 


\begin{table}[h!b!p!]
\begin{tabular}{|c|c|c|c||c|c||c|c|}
\hline
&&\multicolumn{2}{c||}{BNS}&\multicolumn{2}{c||}{BHNS}&\multicolumn{2}{c|}{BBH}\\
&& $\sigma[1]$ & $\sigma[2]$ & $\sigma[1]$ & $\sigma[2]$& $\sigma[1]$ & $\sigma[2]$\\
 \hline
\multirow{7}{3mm}{\begin{sideways}\parbox{15mm}{HLVA}\end{sideways}}&$\mathcal{M}$ [\%]    & 4.56e-3& 2.92e-3 & 2.79e-2 & 1.86e-2 &0.21 & 0.15\\ \cline{2-8}
&$\eta$ [\%]           & 0.22 &0.14      & 0.38 & 0.22 & 1.58  & 0.99 \\ \cline{2-8}
&$t_a$ [ms]            & 0.11 & 0.11     & 0.17& 0.12 & 0.36& 0.24\\ \cline{2-8}
&lat [mrad]             & 9.58 & 9.51    &13.5 & 9.78 & 17.8 & 10.1\\\cline{2-8}
&long [mrad]            & 39.0 & 40.0    &46.6 &31.7 &61.3 & 41.5\\ \cline{2-8}
&$\Delta\Omega [deg^2]$& 1.34& 4.22*  &2.32 &5.11* &4.01 &7.20*\\\cline{2-8}
&Net. SNR & \multicolumn{2}{c||}{15.7} & \multicolumn{2}{c||}{14.9} & \multicolumn{2}{c|}{14.29}\\
\hline
 \hline
\multirow{7}{3mm}{\begin{sideways}\parbox{15mm}{HLVI}\end{sideways}}&$\mathcal{M}$ [\%]      & 4.63e-3& 3.00e-3 & 2.81e-2 & 1.88e-2 &0.21 & 0.15\\\cline{2-8}
&$\eta$ [\%]             & 0.23 &0.16     & 0.39 & 0.23 & 1.61  & 1.03 \\ \cline{2-8}
&$t_a$ [ms]              & 0.17 & 0.17    & 0.27& 0.18 & 0.47& 0.31\\ \cline{2-8}
&lat [mrad]               & 14.6 & 15.3 &20.4 & 14.1 &26.3 & 15.0\\ \cline{2-8}
&long [mrad]              & 46.0 & 48.3 &51.2 &34.6 &66.9 & 47.4\\ \cline{2-8}
&$\Delta\Omega [deg^2]$ & 1.85& 9.57*  &3.09   &8.50* &5.27&10.12*\\ \cline{2-8}
&Net. SNR & \multicolumn{2}{c||}{15.4} & \multicolumn{2}{c||}{14.7} & \multicolumn{2}{c|}{14.05}\\
\hline
 \hline
\multirow{7}{3mm}{\begin{sideways}\parbox{15mm}{HLVJ}\end{sideways}}&$\mathcal{M}$ [\%]     & 4.39e-3& 2.64e-3 & 2.67e-2 &1.68e-2 &0.20 & 0.13\\ \cline{2-8}
&$\eta$ [\%]            & 0.21 & 0.13    & 0.37 & 0.20 & 1.50  & 0.89 \\ \cline{2-8}
&$t_a$ [ms]             & 0.17 & 0.15     & 0.26& 0.17 & 0.44& 0.27\\ \cline{2-8}
&lat [mrad]              & 13.6 & 16.3  &19.5 & 16.1 &25.0 & 17.0\\ \cline{2-8}
&long [mrad]             & 37.4 & 39.3  &44.9 &31.7 &58.8 & 40.7\\ \cline{2-8}
&$\Delta\Omega [deg^2]$ & 1.98& 13.39*    & 3.49 &10.29*  &5.94&12.68*\\ \cline{2-8}
&Net. SNR & \multicolumn{2}{c||}{16.01} & \multicolumn{2}{c||}{15.23} & \multicolumn{2}{c|}{14.57}\\
\hline
\end{tabular}
\caption{ Same as table \ref{Table5Dim3Det} but using 4-IFO networks}\label{Table5Dim4Det}
\end{table}

The same trends are visible in the two five detector networks we have considered: HLVJA and HLVJI.
The results are summarize in table \ref{Table5Dim5Det}. Here again it is visible how the intrinsic parameters do not take any advantage from the addition of a new IFO to the network, keeping a comparable SNR.

\begin{figure}[htb]
\includegraphics[scale=\scale]{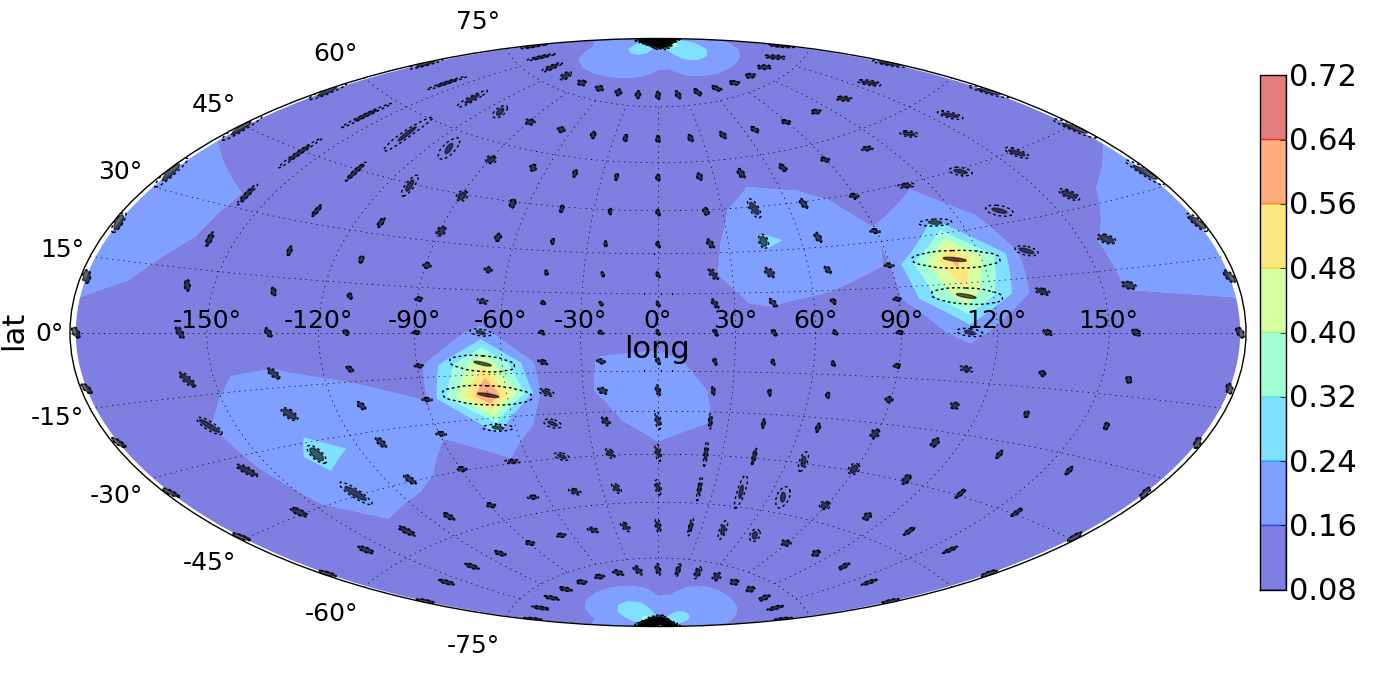}
\caption{(Color online) Skymap for a BNS signal detected in HLVJA. The error on the arrival time is plotted in the background and its value is shown in the colorbar (in ms)}
\label{SkyHLVJA_BNS}
\end{figure}

The angular errors quoted in table \ref{Table5Dim5Det} seem to suggest that HLVJI is the best 5 IFOs network. However one has to remember that what is given in table \ref{Table5Dim5Det} is an average of $\Delta \Omega$ over the 189 sky positions we considered. The reason why the first + second order error $\Delta \Omega[1+2]$ is slightly larger for HLVJA is that there are a few spots in the sky (four, corresponding to positions where the time-resolution is poor) with very large sky errors, $\sim 60\mbox{ deg}^2$, that increase the sky-average. These spots are clearly visible in Fig. \ref{SkyHLVJA_BNS}. 
Aside from these points, HLVJA is better performing. That is seen in table \ref{TableSummary}, in which the number of signals with sky error smaller than 1, 2 or 3 square degrees are given.

It is worth mentioning that this trends in the sky localization accuracy, while adding IFOs to the network,
was investigated with Monte Carlo simulations for burst signals in \cite{Klimenko2011}.
Explicitly it was noted that going from a 4-IFO to a 5-IFO network, the gain in sky localization is 
smaller than going from a 3-IFO to a 4-IFO network. 

In this paper we also observe that the increment in performance 
depend on \emph{which} detectors have been added, and in which order. For example, using only the CRLB only and going from HLV to HLVJ to HLVJA, comparing tables \ref{Table5Dim3Det}, \ref{Table5Dim4Det} and \ref{Table5Dim5Det} it can be seen that the first order $\Delta \Omega[1]$ for BNS systems varies in the following way (in parenthesis, the variation with respect to HLV):

\begin{itemize}
\item HLV: 4.71 $\mbox{deg}^2$
\item HLVJ: 1.98 $\mbox{deg}^2$ (-58\%)
\item HLVJA: 1.02 $\mbox{deg}^2$ (-78\%)
\end{itemize}

On the other hand if the order is HLV, HLVA, HLVJA we find:

\begin{itemize}
\item HLV: 4.71 $\mbox{deg}^2$
\item HLVA: 1.34 $\mbox{deg}^2$ (-72\%)
\item HLVJA: 1.02 $\mbox{deg}^2$ (-78\%)
\end{itemize}

The situation is even more clear when the second order is taken into account. For example if LCGT is the first IFO to be added to HLV the errors $\Delta\Omega[1+2]$ vary as:

\begin{itemize}
\item HLV: 16.4 $\mbox{deg}^2$
\item HLVJ:13.4  $\mbox{deg}^2$ (-18\%)
\item HLVJA: 2.78 $\mbox{deg}^2$ (-83\%)
\end{itemize}

but if the order is HLV, HLVA, HLVJA we find

\begin{itemize}
\item HLV: 16.4 $\mbox{deg}^2$
\item HLVA: 4.22  $\mbox{deg}^2$ (-74\%)
\item  HLVJA: 2.78 $\mbox{deg}^2$ (-83\%)
\end{itemize}

So that if LCGT is the first IFO to be added to HLV, there is still another $20\%$ to be gained at the first order, and $65\%$ using first and second order, by adding AIGO (with Indigo the differences are much smaller, to the close position of LCGT and Indigo on the Earth).

\begin{figure}[htb]
\includegraphics[scale=\scale]{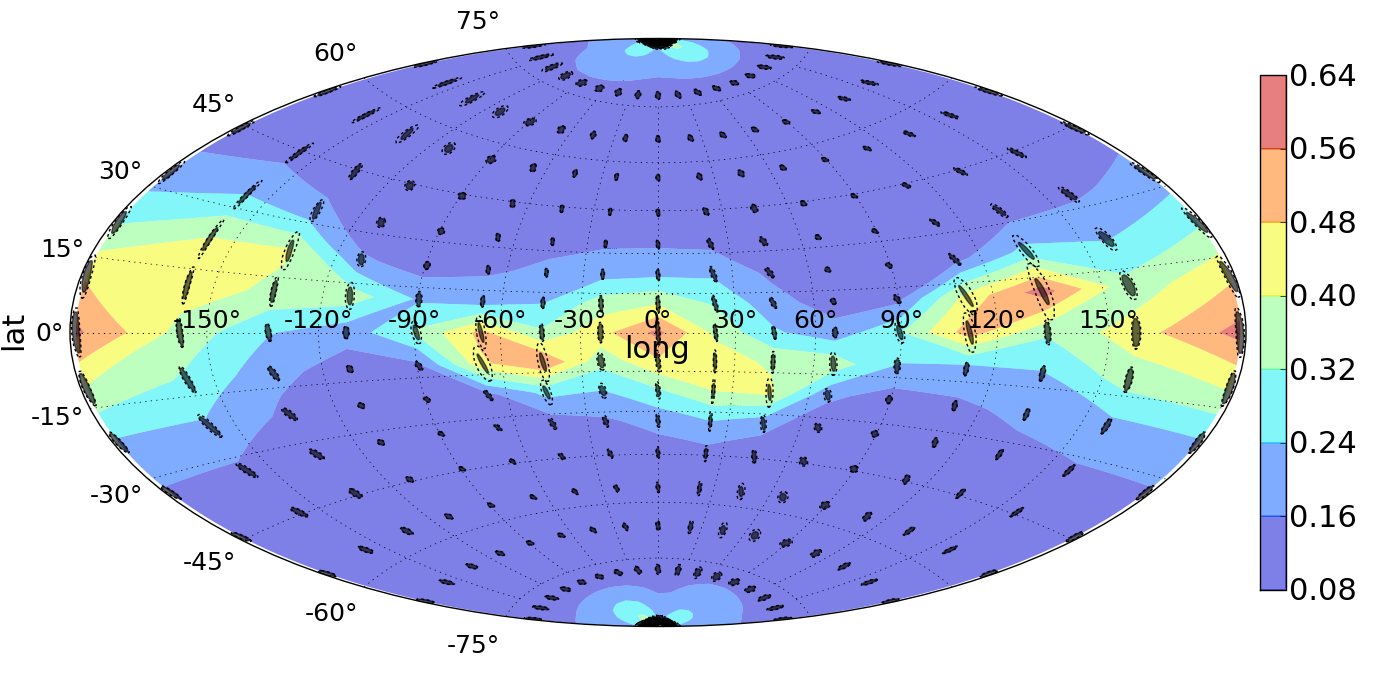}
\caption{(Color online) Skymap for a BNS signal detected in HLVJI. The error on the arrival time is plotted in the background and its value is shown in the colorbar (in ms)}
\label{SkyHLVJI_BNS}
\end{figure}

Note also that the total gain going from 3 to 5 IFOs is not too different using the first order errors only ($-71\%$ with HLVJI and $-78\%$ with HLVJA), or the first and second ($-84\%$ with HLVJI and $-83\%$ with HLVJA).
On fig. \ref{SkyHLVJA_BNS} and \ref{SkyHLVJI_BNS} we plot the sky errors together with the errors on the arrival time.


 \begin{table}[h]
\begin{tabular}{|c|c|c|c||c|c||c|c|}
\hline
&&\multicolumn{2}{c||}{BNS}&\multicolumn{2}{c||}{BHNS}&\multicolumn{2}{c|}{BBH}\\
&& $\sigma[1]$ & $\sigma[2]$ & $\sigma[1]$ & $\sigma[2]$& $\sigma[1]$ & $\sigma[2]$\\
 \hline
\multirow{7}{3mm}{\begin{sideways}\parbox{15mm}{HLVJI}\end{sideways}}&$\mathcal{M}$ [\%]    & 4.47e-3& 2.75e-3 & 2.75e-2 &1.79e-2 &0.21   & 0.15\\ \hline
&$\eta$ [\%]           & 0.21   &0.13    & 0.37  & 0.21& 1.56  &0.97 \\ \cline{2-8}
&$t_a$ [ms]            & 0.14   & 0.11   & 0.22& 0.12   & 0.41  & 0.26\\ \cline{2-8}
&lat [mrad]             & 11.6  & 9.39  &15.2 & 7.68 &20.0  & 9.27\\ \cline{2-8}
&long [mrad]            & 37.3  & 35.6  &42.5 &26.6  &56.9 & 38.8\\ \cline{2-8}
&$\Delta\Omega [deg^2]$& 1.37  &  2.60*   &2.14  & 2.93*  &  3.77  &4.88*\\ \cline{2-8}
&Net. SNR & \multicolumn{2}{c||}{15.93} & \multicolumn{2}{c||}{14.95} & \multicolumn{2}{c|}{14.06}\\
\hline
\hline
&&\multicolumn{2}{c||}{BNS}&\multicolumn{2}{c||}{BHNS}&\multicolumn{2}{c|}{BBH}\\
&& $\sigma[1]$ & $\sigma[2]$ & $\sigma[1]$ & $\sigma[2]$& $\sigma[1]$ & $\sigma[2]$\\
 \hline
\multirow{7}{3mm}{\begin{sideways}\parbox{15mm}{HLVJA}\end{sideways}}&$\mathcal{M}$ [\%] & 4.44e-3& 2.74e-3 & 2.74e-2 & 1.79e-2 &0.21 & 0.15\\ \hline
&$\eta$ [\%]        & 0.21 & 0.14 & 0.37 & 0.21 & 1.55  & 0.96 \\ \cline{2-8}
&$t_a$ [ms]         & 0.11 & 0.10 & 0.17 &0.10 & 0.35& 0.23\\ \cline{2-8}
&lat [mrad]          & 8.19 & 7.47 &11.2 & 7.32 &15.0 & 8.51\\\cline{2-8}
&long [mrad]         & 37.1 & 38.2 &44.0 &29.2 &58.6 & 39.9\\ \cline{2-8}
&$\Delta\Omega [deg^2]$ & 1.02& 2.78*&1.72 &3.17*    &3.09&4.86*\\ \cline{2-8}
&Net. SNR & \multicolumn{2}{c||}{16.18} & \multicolumn{2}{c||}{15.18} & \multicolumn{2}{c|}{14.29}\\
\hline
\end{tabular}
\caption{Same as table \ref{Table5Dim3Det} but using  5-IFO networks}\label{Table5Dim5Det}
\end{table}
(b)
The results of this section suggest a word of caution in reducing the parameter 
space and relying only on the IFIM.

If someone would use the 
two parameter approximation and only the Fisher information matrix,
the square degree uncertainty would be underestimated (on average
for the examples considered in this paper) by a factor 6.7 .

The results of the 2 dimensional parameter space presented in 
tables  \ref{Table2Dim3Det}, \ref{Table2Dim4Det} and \ref{Table2Dim5Det})
are less relevant in the quantitative discussion of the errors 
because the angular resolution is 4.5 times  smaller 
than the values in the tables for the 5 dimensional parameter space presented here.
(notice that the increase in the uncertainties with the size of the 
parameter space was already discussed before for a single IFO parameter estimation 
scenario based on the IFIM, e.g. in \cite{Arun2005}). 

It is also important to use the results including the second order of the expansions.
Such role, with the 5 dimensional parameter is illustrated in the tables V, VI, and VII.
If we include the second order for an average network SNR of 15, 
the solid angle uncertainty increases on average $162\%$, or 
between $79\%$ and $248\%$ with 3 IFOs,
between $80\%$ and $576\%$ with 4 IFOs and
between $29\%$ and $173\%$ with 5 IFOs, depending on the Network and mass bin considered.

\section{Conclusions}

In this paper we have described the most accurate analytical frequentist assessment to date of the uncertainties in the estimation of physical parameters from gravitational waves generated by non spinning binary systems detected by Earth-based networks of laser interferometers. The technique we adopted to quantify the errors is an asymptotic expansion of the uncertainties in inverse powers of the SNR ratio, where the first order is the inverse Fisher information matrix, and it provides results which are better suited to describes low SNR regimes.

We have quantified how the accuracy for the intrinsic parameters  depends on the network SNR, while the measure of the direction of arrival also depends on the network geometry.
We have compared results for 6 different existing and possible global networks and quantified the relative advantages of different proposed sites. In particular, the fraction of the sky where the one sigma angular resolution is below 2 square degrees is shown to increase about 2.3 times when transitioning from the HLV network to a possible 5 sites one (keeping the network SNR fixed). Between the proposed four IFOs networks, HLVA guarantees smaller errors on the reconstruction of the position of the source. On the other hand, the two five IFOs network we considered, HLVJA and HLVJI, seem to be have comparable performances (see \ref{TableSummary}).

\section{Acknowledgments} 
The authors would like to thank P. Ajith and R. Frey for useful comments on the draft of the paper and the LPTMC laboratory of the University UPMC (France) for the use of computers.
S.V. is supported by the research programme of the Foundation for Fundamental Research on Matter (FOM), which is partially supported by the Netherlands Organisation for Scientific Research (NWO). 
S.V. would like to thank W. Del Pozzo and C. Van Den Broeck for useful discussions.
M.Z. thanks the National Science Fondation for the support through the awards NSF855567 and NSF0919034.
\bwt
\begin{table}[htb]
\begin{tabular}{|c|c||c|c|c||c|c|c||c|c|c|}
\hline
&&\multicolumn{3}{c||}{BNS}&\multicolumn{3}{c||}{BHNS}&\multicolumn{3}{c|}{BBH}\\
&& 1 & 2  & 3& 1 & 2  & 3& 1 & 2  & 3 \\
 \hline
\multirow{4}{3mm}{\begin{sideways}\parbox{15mm}{HLVJI}\end{sideways}}&$CL_{68}$[1] [\%]  & 49.48   &  79.24  & 92.39   &18.34 &60.55 & 80.27& 0.0&25.60 &48.10\\ \cline{2-11}
&$CL_{95}$[1] [\%] & 0.0  &  4.15 & 32.53 &0.0&0.0&6.57& 0.0& 0.0&0.0\\ \cline{2-11}
&$CL_{68}$[1+2] [\%] &21.11 & 61.59 & 76.12 &5.88&48.10&69.90 & 0.0&11.07 &42.56  \\ \cline{2-11}
&$CL_{95}$[1+2] [\%]   &0.0  & 0.0  & 3.46 &0.0&0.0&0.0& 0.0&0.0 &0.0\\ \cline{2-11}
\hline\hline
\multirow{4}{3mm}{\begin{sideways}\parbox{15mm}{HLVJA}\end{sideways}}&$CL_{68}$[1] [\%]  &62.63   & 96.19  & 98.61   & 13.84  &69.55 &93.77&    4.84  &21.45 &59.86\\ \cline{2-11}
&$CL_{95}$[1] [\%]  & 0.0  &  5.89 & 31.83 & 0.0  &1.38&8.30& 0.0   & 0.0&0.0\\ \cline{2-11}
&$CL_{68}$[1+2] [\%] &19.72 & 64.71 & 80.28 &7.61  &52.60&67.47& 0.0  &10.38 &43.94  \\ \cline{2-11}
&$CL_{95}$[1+2] [\%]  &0.0  & 0.0  & 3.46 & 0.0&0.0&3.80& 0.0 &0.0 &0.0\\ \cline{2-11}
\hline\hline
 \multirow{4}{3mm}{\begin{sideways}\parbox{15mm}{HLVJ}\end{sideways}}&$CL_{68}$[1] [\%]  &40.83   & 66.44  & 80.62   & 11.76  &41.18 & 57.09&   1.73  &11.76 &35.64\\ \cline{2-11}
&$CL_{95}$[1] [\%]  & 0.0  &  5.88 & 28.72 & 0.0  &0.00&6.92& 0.0   & 0.0&0.0\\ \cline{2-11}
&$CL_{68}$[1+2] [\%] &15.22 & 46.02 & 57.43 &6.92  &24.57&40.83& 0.0  &6.92 &22.15  \\ \cline{2-11}
&$CL_{95}$[1+2] [\%]  &0.0  & 0.0  & 3.46 & 0.0&0.0&2.08& 0.0 &0.0 &0.0\\ \cline{2-11}
\hline\hline
\multirow{4}{3mm}{\begin{sideways}\parbox{15mm}{HLVA}\end{sideways}}&$CL_{68}$[1] [\%]  &45.32   & 87.54  & 96.54   &   11.76   &59.52 & 81.66&   4.15   &11.76 &43.25\\ \cline{2-11}
&$CL_{95}$[1] [\%]  & 0.0  &  5.54 & 27.68 & 0.0  &1.38&7.96 &0.0   & 0.0&0.0\\ \cline{2-11}
&$CL_{68}$[1+2] [\%] &13.49 & 56.05 & 72.32 & 6.92 &33.22&55.36&   0.0&7.61 &23.86  \\ \cline{2-11}
&$CL_{95}$[1+2] [\%]  &0.0  & 0.0  & 4.15 & 0.0&0.0&4.15& 0.0 &0.0 &0.0\\ \cline{2-11}
\hline\hline
\multirow{4}{3mm}{\begin{sideways}\parbox{15mm}{HLVI}\end{sideways}}&$CL_{68}$[1] [\%]  &22.84   & 63.32  & 85.81   & 5.88  &34.60 & 60.55&   0.0  &8.30 &24.91\\ \cline{2-11}
&$CL_{95}$[1] [\%]  & 0.0  &  0.0 & 5.54 & 0.0  &0.0&0.0& 0.0   & 0.0&0.0\\ \cline{2-11}
&$CL_{68}$[1+2] [\%] &0.69 & 36.33 & 57.44 &0.0  &17.65&40.83& 0.0  &0.69 &9.00  \\ \cline{2-11}
&$CL_{95}$[1+2] [\%]  &0.0  & 0.0  & 0.0 & 0.0&0.0&0.0& 0.0 &0.0 &0.0\\ \cline{2-11}
\hline\hline
\multirow{4}{3mm}{\begin{sideways}\parbox{15mm}{HLV}\end{sideways}}&$CL_{68}$[1] [\%]  &13.15   & 42.56  & 54.32   & 1.38  &11.76 & 27.68&   0.0  &4.15 &11.76\\ \cline{2-11}
&$CL_{95}$[1] [\%]  & 0.0  &  0.0 & 6.22 & 0.0  &0.0&0.0& 0.0   & 0.0&0.0\\ \cline{2-11}
&$CL_{68}$[1+2] [\%] &0.0 & 25.26 & 41.18 &0.0  &9.69&20.07& 0.0  &0.0 &4.15  \\ \cline{2-11}
&$CL_{95}$[1+2] [\%]  &0.0  & 0.0  & 0.0 & 0.0&0.0&0.0& 0.0 &0.0 &0.0\\ \cline{2-11}
\hline
\end{tabular}
\caption{Fraction of sources whose confidence levels of 68\% (1 standard deviation) or 95\% (2 standard deviations) are smaller than 1, 2 or 3 square degrees, considering only the first order errors, or the first and second order.}\label{TableSummary}
\end{table}
\ewt

\appendix
\section{Symmetric and trace free tensors and Gel'fand Functions}\label{Appendix_Gelfand}

In this appendix we provide some useful expressions and derivations that are used in the text, advising the reader to refer to \cite{Gelfand}, \cite{Thorne1980} and \cite{Goldber1967} for a complete analysis of the topic. Useful information can also be found in \cite{Maggiore}.

Let us consider a Cartesian frame, call it 'WF', with coordinates $(x,y,z)$. We have seen before that the wave tensor can be written:

\beq\label{w_appendix}
w_{ij}= h_+ \Re[m_i m_j] + h_\times \Im[m_i m_j]
\eeq

where m is complex unit vector, whose components in WF are $\bm{m}= \frac{1}{\sqrt{2}} (1,i,0)$.

Obviously:

\begin{eqnarray}
\Re[m_i m_j] &=& \frac{1}{2} (m_i m_j + m_i^* m_j^*) \label{remm}\\
\Im[m_i m_j] &=& \frac{1}{2\,i} (m_i m_j - m_i^* m_j^*) \label{immm}\\
\end{eqnarray}

The explicit expression of $m \otimes m$ in the WF is:

\begin{equation}\label{mtimesm}
m_i m_j= \frac{1}{2} \left( \begin{array}{ccc}
1 &i&0\\
i& -1 &0\\
0&0&0
  \end{array}\right)
\end{equation}

while $m^* \otimes m^*$ is just the complex conjugate.
Both the matrices are symmetric and trace free (STF), and is then possible to expand them in a base of STF tensor. 
Given a STF tensor, $\bm{Q}$, whose Cartesian coordinates, $Q_{ij}$, are given in the WF, one can expand it in the STF base as follows:

\beq\label{Q}
Q_{ij} = Q_m \mathcal{Y}^{2m}_{ij}
\eeq

where we use the Einstein convention on the repeated indexes, with $m= \pm 2,\pm 1, 0$.
The quantities $Q_m$ are called the spherical components of $Q_{ij}$.

The base tensors $\mathcal{Y}^{2m}_{ij}$ can be put in relation with the usual spherical harmonics $Y^{lm}(\phi,\theta)$ with $l=2$ using the following formula:

\begin{equation}\label{ST}
Y^{2m}(\phi,\theta) = \mathcal{Y}^m_{ij} n^i n^j
\end{equation}

where $n^i$ are the components of the unit radial vector $\bm{n}=(\sin\theta \cos\phi,\sin\theta \sin\phi, \cos\theta)$.
Eq. \ref{ST} can be inverted to find the explicit expressions of the base tensors:

\bwt
\begin{equation}\label{STF_base_Appendix}
 \begin{array}{cc}
 {\mathcal{Y}}^2 \equiv \frac{1}{4}\sqrt{\frac{15}{2\pi}} \left(\begin{array}{ccc} 1&i&0\\ i&-1&0\\ 0&0&0\end{array}\right)\;,& \mathcal{Y}^1 \equiv -\frac{1}{4}\sqrt{\frac{15}{2\pi}} \left(\begin{array}{ccc} 0&0&1\\ 0&0&i\\ 1&i&0\end{array}\right)\;,\\
\mathcal{Y}^0 \equiv \frac{1}{2}\sqrt{\frac{5}{4\pi}} \left(\begin{array}{ccc} -1&0&0\\ 0&-1&0\\ 0&0&2\end{array}\right)\,,& \mathcal{Y}^{-2} = {\mathcal{Y}^{2}}^{*}\;,\; \mathcal{Y}^{-1}= - {\mathcal{Y}^{1}}^*\;
 \end{array}
\end{equation}
\end{widetext}

It is easy to verify that the $\mathcal{Y}$s also satisfy the following closure relation:

\begin{equation}
\mbox{Tr}\left[\mathcal{Y}^{m} {\mathcal{Y}^{n}}^* \right] = \frac{15}{8\pi} \delta_{m\,n}
\end{equation}

and that they satisfy:

\begin{equation}
{\mathcal{Y}^n}^* = (-)^n \mathcal{Y}^{-n}
\end{equation}

which we used already in eq. (\ref{STF_base_Appendix}) to save space.

Eq. \ref{Q} can be inverted to find the spherical components of the tensor Q:

\beq\label{Q_spherical}
Q_m = \frac{8\pi}{15} Q_{ij} {\mathcal{Y}^m_{ij}}^*
\eeq
Formally we can expand  $m \otimes m$ and  $m^* \otimes m^*$  in this base:

\begin{eqnarray}
(m \otimes m)_{ij} &=& M_m \mathcal{Y}^m_{ij} \label{m_expanded}\\
(m^* \otimes m^*)_{ij} &=& P_m {\mathcal{Y}^m_{ij}} \label{mconj_expanded}
\end{eqnarray}

Combining eqs. \ref{w_appendix},\ref{remm},\ref{immm},\ref{m_expanded} and \ref{mconj_expanded} we get:

\beq\label{w_MP}
w_{ij}= \mathcal{Y}^{n}_{ij} \left( \frac{h_+}{2} (M_n + P_n)  +\frac{ h_\times}{2\,i} (M_n - P_n)\right)
\eeq

where the only non-zero components of $M_m$ and $P_m$ are:

\beq
 M_2 = P_{-2}= 2 \sqrt{\frac{2\pi}{15}} 
 \eeq

so that one recovers the expected result from \ref{w_MP}:
\bwt
\begin{eqnarray}
 w_{ij} &=& \mathcal{Y}^{2}_{ij} \left( \frac{h_+}{2} M_2  - i \frac{ h_\times}{2} M_2\right) +\mathcal{Y}^{-2}_{ij} \left( \frac{h_+}{2} P_{-2}  + i \frac{ h_\times}{2} P_{-2}\right) = \nonumber\\
&=&\frac{1}{4} (h_+ - i h_\times) \left(\begin{array}{ccc} 1&i&0\\ i&-1&0\\ 0&0&0\end{array}\right) +
\frac{1}{4} (h_+ +i h_\times) \left(\begin{array}{ccc} 1&-i&0\\ -i&-1&0\\ 0&0&0\end{array}\right)= \nonumber\\
&=&\frac{1}{2} \left(\begin{array}{ccc} h_+&h_\times&0\\ h_\times&-h_+&0\\ 0&0&0\end{array}\right)
\end{eqnarray}
\ewt

Let us now introduce a second frame, called  EF. We are interested in the relation between the components of a STF tensor in the frame WF and in EF.

We assume that the WF is obtained from the EF with a rotation, parametrized with Euler angles $(\alpha,\beta,\gamma)$. The functional dependence on a set of Euler angle is indicated either by writing explicitly the Euler angles or with an arrow pointing from the fix frame to the rotated one. For example:

\beq
F(\alpha,\beta,\gamma)
\eeq
and
\beq
F(EF\rightarrow WF)
\eeq

both indicate a function $F$ depending on the Euler angles that rotate EF ro WF.

Eq. \ref{w_appendix} can still be used, providing that the components of $\bm{m}$ in the EF are used. These are obviously related with the components in the WF:

\beq
{\,}^{EF} \bm{m} = R^{-1}(\alpha,\beta,\gamma)\;{\,}^{WF} \bm{m}
\eeq

where $R^{-1}(\alpha,\beta,\gamma)$ is the inverse rotation $3\times 3$ matrix.

Moreover, we can expand $(m \otimes m)$ and  $(m^* \otimes m^*)$, which are still STF:
\begin{eqnarray}
(m \otimes m)_{ij} &=& \sqrt{\frac{8\pi}{15}} T_{2m}(\mbox{EF}\rightarrow \mbox{WF}) \mathcal{Y}^m_{ij} \label{mmEF}\\
(m^* \otimes m^*)_{ij} &=& \sqrt{\frac{8\pi}{15}} T_{-2\,m}(\mbox{EF}\rightarrow \mbox{WF}) \mathcal{Y}^m_{ij} \label{mcmcEF}
\end{eqnarray}

where we have used $T_{ a b}$ to denote the \emph{Gel'fand functions} of rank 2, which are nothing else than the spherical components of $(m \otimes m)$ and  $(m^* \otimes m^*)$ in this rotated frame.
We have explicitly written down the dependence of $T_{\pm2 n}$ on the rotation matrix (i.e. on the angles that parametrize the rotation).

The previous relations can be inverted, to find the explicit expressions of the Gel'fand functions:

\begin{eqnarray}
T_{2\,m}(\mbox{EF}\rightarrow \mbox{WF}) &=& \sqrt{\frac{8\pi}{15}} {\mathcal{Y}^m_{ij} }^* m_i m_{j}\\
T_{-2\,m}(\mbox{EF}\rightarrow \mbox{WF}) &=& \sqrt{\frac{8\pi}{15}}  {\mathcal{Y}^m_{ij} }^* m_i^* m_{j}^*
\end{eqnarray}

Some of the symmetries own by the rank-2 Gel'fand functions are easier to prove starting from the most general definition, of the rank l functions:

\begin{equation}\label{Gelfand_generalized}
 T^l_{mn} (\alpha,\beta,\gamma) \equiv e^{-i m \gamma} P^l_{mn}(\cos(\beta)) e^{-i n \alpha}
\end{equation}

from which the ones we used are obtained putting $l=2$ (we shall drop out the apex 2, as we only use the rank-2 functions so no confusion is possible).
In eq. \ref{Gelfand_generalized} we have introduced the function\footnote{Note that l - n=0 does {\bf not} imply that the term within derivative does not give contribution, but only that the derivative is not to be performed: $\frac{d^0 A(\mu)}{d\mu^0}\equiv A(\mu)$}:
\bwt
\begin{equation}
 P^l_{mn}(\mu) = A (1-\mu)^{-\frac{n-m}{2}} (1+\mu)^{-\frac{n+m}{2}} \frac{d^{l-n}}{d\mu^{l-n}} \left[(1-\mu)^{l-m} (1+\mu)^{l+m}\right] \;\; 
\end{equation}
\ewt
where the numerical coefficient $A$ has the value

\begin{equation}
 A\equiv \frac{(-1)^{l-m} i^{n-m} }{2^l (l-m)!} \sqrt{\frac{(l-m)!\,(l+n)!}{(l+m)!\,(l-n)!}} 
\end{equation}

Is is easy to check that $P^l_{mn}$ has the following symmetries:

\begin{equation}
P^l_{-n\,-m}= P^l_{nm}=P^l_{mn}= (-1)^{m+n}\,{P^l_{mn}}^{*}.
\end{equation}

that can be used to show that:

\begin{equation}\label{simmetry_T}
 T_{-m\,-n}= (-)^{-m-n} T_{m n}^{\,*}
\end{equation}

Finally it can be shown (\cite{Gelfand}) that there is a very simple relation between the  Gel'fand function associated with a rotation $R$ and that associated with the inverse rotation:

\begin{equation}\label{Gelfand_inverse}
 T_{m\,n}(R^{-1}) =T^{-1}_{m\,n}(R) = T^*_{n\,m}(R).
\end{equation}

Using eq. \ref{mmEF} and \ref{mcmcEF} on eq. \ref{w_appendix}, the wave tensor in the EF can be written:

\beq\label{w_general}
w_{ij}= \mathcal{Y}_{ij}^n \sqrt{\frac{2\pi}{15}} \left( h_+ ( T_{2n} + T_{-2n}) + i h_\times  ( T_{2n} - T_{-2n}) \right)
\eeq

where it is understood that the $T_{\pm 2 n} = T_{\pm 2 n} (\mbox{EF}\rightarrow \mbox{WF}) = T_{\pm 2 n} ( \alpha,\beta,\gamma)$.

One of the advantages of having a STF tensor expanded in the STF base, other than the geometrical insight, is that one can take advantages of all the proprieties of the
Gel'fand function.
In the main text, for example, we need to write the detector tensor of the I-th IFO (which is also a STF tensor) in the wave frame.  We can apply to that tensor the same manipulations described in this appendix, and obtain an expression containing: $ T_{\pm 2 n} (\mbox{WF} \rightarrow \mbox{IDT})$ where we need the angles that rotate the Wave frame in the I-th detector frame. If many detectors are present in the network, however, it would be better to calculate the position of source in a common frame (using eq. \ref{Euler_to_Spherical} the Euler angles can be obtained), which can be the Earth frame. 
We can achieve that result, writing the rotation $\mbox{WF} \rightarrow \mbox{IDT}$ in terms of the two successive rotations $\mbox{WF}\rightarrow \mbox{EF}$ and $\mbox{EF}\rightarrow \mbox{IDF}$. The behavior of the Gel'fand functions over two successive rotations in very simple. 
Let $R_1$ and $R_2$ be two rotations, then the Gel'fand function associated with $R\equiv R_{2}\,R_1$ can be obtained from those associated with $R_1$ and $R_2$ as follows:

\begin{equation}\label{Gelfand_sum}
T_{mn}(R_2\,R_1)= \sum^{2}_{s=-2}{ T_{ms}(R_2)\,T_{sn}(R_1)}
\end{equation}

which allows us to write:
\begin{eqnarray}
T_{\pm 2 n} (\mbox{WF} \rightarrow \mbox{IDT}) =T_{\pm 2 n} (\mbox{EF}\rightarrow \mbox{IDF},\mbox{WF}\rightarrow \mbox{EF}) &=& \nonumber \\
=\sum^{2}_{s=-2} { T_{\pm2 s}(\mbox{EF}\rightarrow \mbox{IDF})\,T_{sn}(\mbox{WF}\rightarrow \mbox{EF})}&&\qquad
\end{eqnarray}

\section{Proof that the generalized antenna pattern go to the single ones}\label{Appendix_Generalized_to_Single}

In this appendix we show that the generalized antenna patterns, eq. \ref{Generalized_antenna_p} and eq. \ref{Generalized_antenna_p} do become equal to eq. \ref{antenna_pattern_p} and eq. \ref{antenna_pattern_p} if a single IFO is present in the network.

If our network is composed of the I-th detector only, we can identify the fide frame with the detector frame, and then it follows from the definition \ref{EF_to_DF} that:

\beq
\alpha^I=\beta^I=\gamma^I=0
\eeq

because now FF$\equiv$IDF.

For the same reason (eq. \ref{time_shift}): $\tau_I=0.$

Let us start from $\chi_2$, eq. \ref{chi_explicit}, in which we now set $\alpha^I=\beta^I=\gamma^I=0$:

\beq \label{chi2_singleIFO}
\chi_2= T^*_{2s}(\phi,\theta,\psi)\left(T_{2s}(0,0,0) + T_{-2s}(0,0,0)\right)
\eeq

Where $\phi,\theta,\psi$ still indicate the Euler angles that rotate the FF (i.e. the IDF) to the WF, eq. \ref{EF_to_WF}.
Starting from the general definition of the Gel'fand functions, eq. \ref{Gelfand_generalized}, is a matter of a few moments proving that:

$$ T_{\pm 2 s}(0,0,0)= \delta_{\pm 2}^s $$

where, as done before, we dropped out the apex, because $l=2$ in this whole article.

Eq. \ref{chi2_singleIFO} then becomes:

\beq \label{chi2_singleIFO_2}
\chi_2= T^*_{2s}(\phi,\theta,\psi)\left(\delta_{2}^s + \delta_{-2}^s\right) =T^*_{22} + T^*_{2-2}.
\eeq

Using again the definition of the Gel'fand function is easy to find the explicit expressions of the two terms above:

\begin{eqnarray*}
T^*_{2\,2}(\phi,\theta,\psi) &=& e^{2 i \psi} e^{2 i \phi} P_{2\,2}^*(\theta)= e^{2 i \psi} e^{2 i \phi} \frac{(1+\cos\theta)^2}{4} \\
T^*_{2\,-2}(\phi,\theta,\psi) &=& e^{2 i \psi} e^{-2 i \phi} P_{2\,-2}^*(\theta)= e^{2 i \psi} e^{-2 i \phi} \frac{(1-\cos\theta)^2}{4} \\
\end{eqnarray*}

that leads to the final expression for $\chi_2$:
\begin{eqnarray*}
\chi_2 &=& \left(\frac{(1+ \cos^2\theta)}{2} \cos 2\psi \cos 2\phi - \cos\theta \sin 2\psi \sin 2\phi \right) +\nonumber\\
+&i& \left(\frac{(1 +\cos^2\theta)}{2} \sin 2\psi \cos 2\phi + \cos\theta \cos 2\psi \sin 2\phi \right).
\end{eqnarray*}

The antenna patterns can be obtained from $\chi_2$ as shown in the main text:

\begin{eqnarray}
\mathcal{F}_+ &=& \Re[\chi_2]\\
\mathcal{F}_\times &=& - \Im[\chi_2]
\end{eqnarray}

from which eqs. \ref{antenna_pattern_p} and \ref{antenna_pattern_c} follow immediately.
\section{Second order non-diagonal elements of the covariance matrix and second order bias}\label{Appendix_bias2}

In this appendix we give the explicit expression of the second order bias and second order non-diagonal elements of the covariance matrix, for a network of N IFOs. 
The $\upsilon$ to be used are the network ones, given in the main text, eqs. \ref{vab-final} $\cdots$ \ref{vabcde-final}.
For the nondiagonal terms of the covariance matrix we find:
\bwt
\begin{eqnarray}
\sigma^2_{\vartheta_i \vartheta_j}[2]&=& \Gamma^{i m} \Gamma^{j n}\Gamma^{p q}\left[ \upsilon_{n m  p q}+3 \sum_{I=1}^N \langle s_{n q} | s_{p m}\rangle + 2\upsilon_{n m p,q}+\frac{1}{2} \upsilon_{m p q,n} +\frac{1}{2} \upsilon_{n p q,m}\right]  + \nonumber\\
&+& \Gamma^{i m}\Gamma^{j n}\Gamma^{p z} \Gamma^{qt}\left[\upsilon_{n p m}\upsilon_{q z t} +\frac{5}{2} \upsilon_{n p q}\upsilon_{m z t} +\upsilon_{q z,n}\upsilon_{m p t}+\upsilon_{qz,m}\upsilon_{n p t} \right.\nonumber\\
 &+& 2\upsilon_{qp,z}\upsilon_{n m t} +3\upsilon_{n t,z}\upsilon_{mpq}+3\upsilon_{m t,z}\upsilon_{n q p}+\frac{1}{2}\upsilon_{m t,n}\upsilon_{p q z}+\frac{1}{2}\upsilon_{n t,m}\upsilon_{p q z}+\nonumber\\
&+& \upsilon_{p t,m}\upsilon_{n q,z}+ \upsilon_{m q,z}\upsilon_{p t,n}+\upsilon_{n q,m}\upsilon_{p t,z}+\upsilon_{m q,n}\upsilon_{p t,z}+\upsilon_{n q,p}\upsilon_{m z,t}\bigg].
\end{eqnarray}
\ewt 
It is easy to show, even though somehow lenghty, that the expression we gave in the text for the diagonal elements, eq. \ref{VarMatrix}, follows from this one setting $i=j$. 
The second order bias has the explicit expression:
\bwt
\begin{eqnarray}\label{IMR-BiasTwoSimplified}
b_{\;\vartheta^m}[2] &=& \frac{\Gamma^{ma}\Gamma^{bd}\Gamma^{ce}}{8}[v_{abcde} + 4 \sum_{I=1}^N\langle s_{ac}\,|\,s_{bde}\rangle^I + 8 \sum_{I=1}^N \langle  s_{de} \,|\,s_{abc}\rangle^I + 4v_{abce,d}] \nonumber\\
&+&\frac{\Gamma^{ma}\Gamma^{bc}\Gamma^{df}\Gamma^{eg}}{4}\bigg[(2v_{afed}v_{gb,c} + 2v_{bedf}v_{ac,g} + 4v_{abed}v_{gf,c}) + (v_{afed}v_{gcb} +\nonumber\\
&+& 2v_{abed}v_{gcf} + 2v_{dbeg}v_{acf}) + ( 2 v_{aed} \sum_{I=1}^N \langle  s_{gb}\,|\,s_{fc}\rangle^I + 4 v_{acf} \sum_{I=1}^N \langle  s_{dg}\,|\,s_{eb}\rangle^I + 4 v_{bed} \sum_{I=1}^N \langle  s_{ac} \,|\,s_{gf}\rangle^I\nonumber  \\
&+&2v_{fcb} \sum_{I=1}^N \langle  s_{ag}\,|\,s_{ed}\rangle^I ) + (4v_{afe,g}v_{db,c} + 4v_{afe,c}v_{db,g} + 4v_{dbe,g}v_{af,c}) + (2v_{abe,g}v_{cdf}\nonumber\\
&+& 4v_{dbe,g}v_{acf} + 4v_{abe,f} v_{cdg} + 2v_{dge,b} v_{acf}) + (4 \sum_{I=1}^N \langle  s_{ag}\,|\,s_{fc}\rangle^I \, v_{ed,b} + 4 \sum_{I=1}^N \langle s_{ed}\,|\,s_{fc}\rangle^I \, v_{ag,b} \nonumber \\
&+& 4 \sum_{I=1}^N \langle  s_{ag}\,|\,s_{ed}\rangle^I \,v_{fc,b}) \bigg] \nonumber \\
&+& \frac{\Gamma^{ma}\Gamma^{bc}\Gamma^{de}\Gamma^{fg}\Gamma^{ti}}{8}[v_{adf}(v_{ebc}v_{gti} + 2v_{etc}v_{gbi} + 4v_{gbe}v_{tci} + 8v_{gbt}v_{eci} + 2v_{ebc}v_{gt,i} \nonumber \\
&+& 4v_{etc}v_{gb,i} + 2v_{gti}v_{eb,c} + 4v_{gtc}v_{eb,i} + 8v_{gbt}v_{ce,i} + 8v_{gbt}v_{ci,e} + 8v_{gbe}v_{ct,i} + 8v_{cte}v_{gb,i} \nonumber \\
&+& 4v_{cti}v_{gb,e} + 4v_{gt,i}v_{eb,c} + 4v_{eb,i}v_{gt,c} + 8v_{gt,b}v_{ic,e} + 8v_{gt,e}v_{ic,b} + 4v_{bet}v_{g,c,i}) \nonumber \\
&+& v_{dci}(8v_{bgt}v_{ae,f} + 4v_{bgf}v_{ae,t} + 8v_{ae,t}v_{bg,f} + 8v_{ae,f}v_{bg,t} + 8v_{af,b}v_{ge,t})]
\end{eqnarray} 
\ewt

\section{Single IFO upsilon}\label{Appendix_upsilon}

In this appendix we give the explicit form of the single IFO terms that appear in the second order covariance and bias. In this appendix it will be assumed that all the quantities refer to one of the IFOs in the network, and the apex $(I)$ to unburden the notation is dropped out.

Let us recall that we have defined:

\begin{eqnarray*}\langle a \cdots i | j \cdots p \rangle &\equiv& \langle s_{a \cdots i} | s_{j \cdots  p} \rangle = \langle \frac{\partial s}{\partial \theta^a \cdots \partial \theta^i} | \frac{\partial s}{\partial \theta^j \cdots \partial \theta^p}  \rangle = \\
&=& 4 \Re \int_{f_{low}}^{f_{cut}}{ \ud f\, \frac{s_{a \cdots i}s_{j \cdots  p}^* }{S(f)}}
\end{eqnarray*}

and that $\Gamma_{i j}$ is the Fisher matrix.

Using this notation, the single IFO $\upsilon$'s can be written (\cite{Vitale2010}):

\begin{eqnarray*}
\upsilon_{a,b}&=& - \upsilon_{ab} = \Gamma_{ab}=\langle a\,|\,b \rangle  \\
\upsilon_{ab\,,\,c} & =& \langle ab\,|\,c\rangle\\
\upsilon_{abc\,,\,d}  &=& \langle abc\,|\,d\rangle \\
\upsilon_{abc} &=& -\langle ab\,|\,c \rangle - \langle ac\,|\,b \rangle-\langle bc\,|\,a \rangle\\
\upsilon_{ab\,,\,cd}&=& \langle ab\,|\,cd \rangle  + \langle a\,|\,b \rangle \langle c\,|\,d \rangle\\
\upsilon_{abcd}&=& -\langle ab\,|\,cd \rangle - \langle ac\,|\,bd \rangle -\langle ad\,|\,bc \rangle -\nonumber \\
&&-\langle abc\,|\,d \rangle-\langle abd\,|\,c \rangle-\langle acd\,|\,b \rangle-\langle bcd\,|\,a \rangle\nonumber \\
\upsilon_{ab\,,\,c\,,\,d}&=& -\langle a\,|\,b \rangle \,\langle c\,|\,d \rangle =-\Gamma_{ab}\,\Gamma_{cd}\\
 \end{eqnarray*}
 \begin{eqnarray*}
\upsilon_{abc\,,\,de} &=& \langle abc\,|\,de\rangle - \Gamma_{de} v_{abc}\\
\upsilon_{abcd\,,\,e} &=& \langle abcd\,|\,e\rangle \\
\upsilon_{abc\,,\,d\,,\,e} &=&  \Gamma_{de} v_{abc}\\
\upsilon_{ab\,,\,cd\,,\,e}&=& - \Gamma_{ab} v_{cd\,,\,e} - \Gamma_{cd} v_{ab\,,\,e}\\
\upsilon_{abcde} &=& -\langle abcd\,|\,e\rangle -\langle abce\,|\,d\rangle-\langle abde\,|\,c\rangle
-\langle acde\,|\,b\rangle \nonumber \\
&&- \langle bcde\,|\,a\rangle -\langle abc\,|\,de\rangle -\langle abd\,|\,ce\rangle
-\langle acd\,|\,be\rangle \nonumber \\
&&- \langle bcd\,|\,ae\rangle -\langle abe\,|\,cd\rangle-\langle ace\,|\,bd\rangle
-\langle bce\,|\,ad\rangle \nonumber \\
&&-\langle ade\,|\,bc\rangle -\langle bce\,|\,ac\rangle -\langle cde\,|\,ba\rangle 
\end{eqnarray*}

\end{document}